\documentclass[aps,pra,superscriptaddress,showpacs,showkeys,a4paper,10pt,notitlepage]{revtex4-1} 
\usepackage[utf8]{inputenc}
\usepackage[english]{babel}
\usepackage{amsmath}
\usepackage{amssymb}
\usepackage{amsfonts}
\usepackage{graphicx}
\usepackage{braket}
\usepackage{bbold}
\usepackage{cancel}
\usepackage{upgreek}
\usepackage{mathtools}
\usepackage[T2A]{fontenc}

\usepackage{xcolor}
\definecolor{link_blue}{RGB}{52,46,157}

\usepackage{longtable}
\usepackage[colorlinks,citecolor=link_blue,linkcolor=link_blue,urlcolor=link_blue]{hyperref}
\usepackage[tiny,center,uppercase]{titlesec}

\setcitestyle{super}

\renewcommand{\vec}{\boldsymbol}

\DeclareMathOperator{\re}{Re}

\DeclareMathOperator{\Tr}{Tr}
\DeclareMathOperator{\Mu}{\mathcal{M}}

\begin{document}

\title{Modelling many-body quantum dynamics with stochastic trajectories: a critical test on the Tavis-Cummings model}

\author{A.\ Leonau}
\email[Corresponding author: ]{aliaksandr.leonau@desy.de}
\affiliation{Deutsches Elektronen-Synchrotron DESY, Hamburg 22607, Germany}

\author{S.\ Chuchurka}
\affiliation{Deutsches Elektronen-Synchrotron DESY, Hamburg 22607, Germany}
\affiliation{Department of Physics, Universit\"at Hamburg, Hamburg 22761, Germany}

\author{V.\ Sukharnikov}
\affiliation{Deutsches Elektronen-Synchrotron DESY, Hamburg 22607, Germany}
\affiliation{Department of Physics, Universit\"at Hamburg, Hamburg 22761, Germany}

\author{A.\ Benediktovitch}
\affiliation{Deutsches Elektronen-Synchrotron DESY, Hamburg 22607, Germany}

\author{N.\ Rohringer}
\email[Corresponding author: ]{nina.rohringer@desy.de}
\affiliation{Deutsches Elektronen-Synchrotron DESY, Hamburg 22607, Germany}
\affiliation{Department of Physics, Universit\"at Hamburg, Hamburg 22761, Germany}

\begin{large}

\begin{abstract}
We critically explore the applicability of a recently proposed framework to sample the quantum dynamics of a many-body quantum system interacting with light by stochastic trajectories, applying it to the closed and open Tavis-Cummings model (TCM).The stochastic differential equations (SDEs) sample the positive P phase-space representation by analog complex-valued dynamical variables that are linked to the quantum operators. Statistical average over the stochastic trajectories yields the evolution of the quantum mechanical expectation values.  However, numerical implementation of these SDEs for the TCM indicates divergent solutions, also known from other phase-space methods. This limits the applicability of the framework to finite propagation times, that are strongly dependent on the physical parameters and initial conditions of the system. We outline the underlying mathematical reason for these divergences and show that their contribution to the averages are, however, essential. To attempt to regularize the divergences, we transform the SDEs to an equivalent set of SDEs with different noise realisations, thereby pushing the valid time boundary. Quantum collapse and revival of the TCM, however, cannot be recovered by the stochastic trajectory approach, pointing to the general difficulty of the applicability of stochastic phase-space sampling methods to systems with strong quantum features.
\end{abstract}

\maketitle



\section{Introduction}
\label{sec:intro}

Solving the quantum equation of motions for  macroscopic ensembles of quantum emitters and a quantized electromagnetic field remains a delicate problem in theoretical physics. To realistically represent experimental systems, the interaction of the quantum system with the environment needs to be considered. The environment is often treated as a reservoir with an infinite number of degrees of freedom \cite{Breuer_open_quantum}. Interaction with the reservoir causes the system's density matrix to undergo decoherence. The dimensionality of the system's Liouville space grows exponentially with respect to the number of emitters, rendering the numerical treatment of these systems computationally intractable for large particle numbers \cite{carmichael_open_1993,chiang_non-markovian_2021,adhikary_dissipative_2023}. To partially overcome this issue, one can impose permutational symmetry of the emitters, thereby reducing the dimension to polynomial scaling \cite{richter_numerically_2015,gegg_efficient_2016,gegg_identical_2017,shammah_open_2018,sukharnikov_second_2023}. By this assumption, the evolution of the many-body density matrix of systems up to the order of 100 emitters can be numerically simulated. For larger system size, alternative approaches based on phase-space sampling methods have been proposed \cite{glauber_coherent_1963,sudarshan_equivalence_1963,wigner_quantum_1932,husimi_formal_1940,drummond_generalised_1980,olsen_phase-space_2005}. These techniques rely on representing the density operator of the quantum system in terms of coherent states and an appropriate phase-space distribution function. Choosing the properties of this function wisely, it can  be interpreted as a quasi-probability density distribution function, and its evolution is governed by a generalized Fokker-Planck equation. The Fokker-Planck equation, in turn, can be mapped to stochastic differential equations (SDEs), providing a pathway for statistical sampling of the system's phase-space evolution \cite{carmichael_statistical_1999,carmichael_statistical_2008} by means of stochastic trajectories of the underlying dynamical variables. One specific example  is the use of the  positive P-representation for the bosonic fields \cite{drummond_generalised_1980,gilchrist_positive_1997,deuar_first-principles_2006}, which has been successfully applied in various problems of quantum optics \cite{carter_squeezing_1987,plimak_optimization_2001,mandt_stochastic_2015,olsen_optical_2018,teh_simulation_2017,wang_fokker-planck_1995,wang_fokker-planck_1996} and Bose-Einstein condensates \cite{drummond_canonical_2004,deuar_correlations_2007,wuster_quantum_2017,deuar_fully_2021}.\\

While the phase-space techniques seem to offer a promising solution for large system sizes, they typically come with their own challenges. One particular issue that arises from the nature of the resulting set of SDEs is the presence of intrinsically divergent stochastic trajectories. These divergences restrict the applicability of the method to short time scales (compared to characteristic times of incorporated quantum effects) and present challenges in obtaining reliable long-term predictions. Dissipation has been recognized as an effective means of regularizing the divergent behavior in certain cases \cite{mandt_stochastic_2015,naether_stationary_2015,deuar_fully_2021}. Another technique known as stochastic drift-gauge transformation \cite{deuar_gauge_2002,drummond_quantum_2003,deuar_thesis_2005,deuar_first-principles_2006-1} has been introduced to address this issue. The strategy behind stochastic gauging is to modify the explicit form of the SDEs while preserving their averages (and thus quantum mechanical expectation values), thereby potentially regulating the divergent behavior of the trajectories.

Here, we critically explore limitations of the stochastic trajectory method by studying the Tavis-Cummings model (TCM) \cite{tavis_exact_1968}. The TCM provides a simplified framework for studying the collective interaction between an ensemble of identical two-level atoms and a "single-mode" quantum field, capturing essential aspects of the light-matter interaction processes. This model has found extensive applications across various physical systems, including cavity quantum electrodynamics (QED) \cite{brennecke_cavity_2007,goryachev_high-cooperativity_2014,wickenbrock_collective_2013,blaha_beyond_2022}, circuit QED \cite{fink_dressed_2009,dicarlo_demonstration_2009,baksic_controlling_2014}, trapped ions \cite{retzker_tavis-cummings_2007,sharma_multipartite_2008,burd_quantum_2021,groszkowski_reservoir-engineered_2022}, and quantum dots \cite{kim_strong_2011,deng_coupling_2015,mohamed_quantum_2019,bhatt_holsteinprimakoff_2023}. Furthermore, we aim to uncover the underlying reasons for the divergent behavior observed in the stochastic trajectories. We investigate potential methods to mitigate the problem of diverging trajectories.

The paper is organized as follows: In Section \ref{sec:2}, we present the Hamiltonian of the TCM and the quantum master equation. We derive the stochastic equations of motion (EOM) in the Ito's form for both the atomic and field degrees of freedom and discuss their relation to the EOM for the averages of the corresponding quantum operators. In Section \ref{sec:3-1}, we present numerical simulations and parameter studies. We investigate regularization by introducing damping to the system. In Section \ref{sec:3-2}, we analyze the nature of the runaway trajectories by examining the semi-classical system of equations and introducing an effective potential for the population inversion. In Section \ref{sec:3-3}, we explore regularization of the diverging stochastic differential equations by reconfiguration of the deterministic terms of the stochastic equations through a procedure known as drift-gauge transformation. Conclusions are summarized in Section \ref{sec:4}. The main text of the paper is supported with Appendices \ref{sec:appA}--\ref{sec:appE}, which provide detailed derivations and additional discussion.

\section{The stochastic equations of motion}
\label{sec:2}

The Hamiltonian of the TCM of a system of $N$ identical two-level atoms interacting with a single-mode quantum field reads \cite{tavis_exact_1968}:
\begin{equation}
	\hat{H} = \frac{\hbar \omega_0}{2} \sum_{\mu = 1}^{N} \Bigl( \hat{\sigma}_{ee}^{(\mu)} - \hat{\sigma}_{gg}^{(\mu)} \Bigr) + \hbar \omega_c \hat{a}^\dagger \hat{a} + \frac{\hbar f}{\sqrt{N}}
		\sum_{\mu = 1}^{N} \Bigl( \hat{\sigma}_{eg}^{(\mu)} \hat{a} + \hat{\sigma}_{ge}^{(\mu)} \hat{a}^\dagger \Bigr),
\label{H_TCM_proj}
\end{equation}
\noindent where $\hat{a}$, $\hat{a}^\dagger$ are the annihilation and creation operators of the quantum field with the frequency $\omega_c$, $\omega_0$ the transition frequency between the atomic levels, $f$ the atom-field coupling constant and $\hbar$ the Planck constant. $\hat{\sigma}^{(\mu)}_{ij}$  are the $\hat{\sigma}$-projectors acting in the Hilbert space of the $\mu$-th atom and defined as
\begin{equation}
	\hat{\sigma}^{(\mu)}_{ij} = \ket{i}_{\mu}\bra{j}_{\mu}, \quad i,j \in \{e,g\}.
\end{equation}
\noindent Here and below the indices ${g}$ and ${e}$ stand for the ground and excited states of the two-level atom, respectively. 

The time evolution of systems's d.o.f is governed by the quantum master equation for its total density operator \cite{manzano_short_2020}. Here, we include dissipative processes, which are often introduced to improve convergence of the stochastic phase-space sampling approaches \cite{mandt_stochastic_2015,naether_stationary_2015,deuar_fully_2021}. To that end, we introduce a Lindblad superoperator treating the decay of the excited states with the rate $\gamma$. We assume that all atoms interact with the reservoir independently but their relaxation rate being equivalent. The corresponding quantum master equation reads:
\begin{equation}
	\frac{d \hat{\rho} (t)}{d t} = \mathcal{L} [\hat{\rho} (t)] = - \frac{i}{\hbar} \left[ \hat{H}, \hat{\rho} (t) \right] + 
		\frac{\gamma}{2} \sum_{\mu = 1}^N \left( 2 \hat{\sigma}^{(\mu)}_{ge} \hat{\rho} (t) \hat{\sigma}^{(\mu)}_{eg} -
		\hat{\sigma}^{(\mu)}_{eg} \hat{\sigma}^{(\mu)}_{ge} \hat{\rho} (t) - \hat{\rho} (t) \hat{\sigma}^{(\mu)}_{eg} \hat{\sigma}^{(\mu)}_{ge}  \right).
	\label{eq:masterdiss}	 	
\end{equation}  

\subsection{The stochastic trajectory formalism}

To highlight the relation between the quantum and stochastic equations of motion, we start our derivation with the EOM for the averages of the quantum field and atomic operators.  We consider full permutational symmetry of the atomic subsystem within the TCM and therefore drop the $\mu$-index. As a result, we obtain \cite{scully_zubairy_1997,benediktovitch_2019}:
\begin{gather}
	\frac{d \langle \hat{a} \rangle}{dt}    = - i \omega_c \langle \hat{a} \rangle - i f \sqrt{N} \langle \hat{\sigma}_{ge} \rangle, \nonumber \\
	\frac{d \langle \hat{a}^\dagger \rangle}{dt}  = ~~i \omega_c \langle \hat{a}^\dagger \rangle + i f\sqrt{N} \langle \hat{\sigma}_{eg} \rangle, \nonumber \\
	\frac{d \langle \hat{\sigma}_{ee} \rangle}{dt}  = - \gamma \langle \hat{\sigma}_{ee} \rangle + i \frac{f}{\sqrt{N}} \Bigl( \langle \hat{\sigma}_{ge} \hat{a}^\dagger \rangle - \langle \hat{\sigma}_{eg} \hat{a} \rangle \Bigr),
	\label{eqn:quantum}\\
	\frac{d \langle \hat{\sigma}_{ge} \rangle}{dt}  = - \frac{\gamma}{2} \langle \hat{\sigma}_{ge} \rangle - i \omega_0 \langle\hat{\sigma}_{ge}\rangle + i \frac{f}{\sqrt{N}} \Bigl( 2 \langle \hat{\sigma}_{ee} \hat{a} \rangle - 
		\langle  \hat{a} \rangle\Bigr), \nonumber\\
	\frac{d \langle \hat{\sigma}_{eg} \rangle}{dt} = - \frac{\gamma}{2} \langle \hat{\sigma}_{eg} \rangle + i \omega_0 \langle\hat{\sigma}_{eg}\rangle - i \frac{f}{\sqrt{N}} \Bigl( 2 \langle \hat{\sigma}_{ee} \hat{a}^\dagger \rangle - 
		\langle \hat{a}^\dagger \rangle\Bigr). \nonumber
\end{gather}
Note that the EOM of the expectation value for the atomic $\sigma$ operators contain the second order atom-field correlation functions and therefore do not represent a closed set of EOMs. Factorization of those second order correlators leads to the semi-classical Maxwell-Bloch equations\cite{dicke_1954,GROSS1982301}. The general idea of the stochastic approach is to introduce additional stochastic terms in the semi-classical Maxwell-Bloch equations that restore the missing correlations (see Appendix \ref{sec:appA-I} for a detailed derivation). A more formal method discussed in Appendix \ref{sec:appA-II} is based on stochastic sampling of the temporal evolution of the positive P-representation of the total density operator \cite{drummond_generalised_1980,gilchrist_positive_1997,deuar_first-principles_2006,chuchurka_quantum_2023}. In reference \cite{chuchurka_quantum_2023} a framework is highlighted, that maps the quantum operators to stochastic phase-space variables 
\begin{gather}
	\hat{a} \rightarrow \alpha (t), \quad \hat{a}^\dagger \rightarrow \alpha^\dagger (t); \quad \hat{\sigma}_{ij} \rightarrow \rho_{ji} (t),
\label{stoch_trans}
\end{gather}
and directly establishes the corresponding stochastic EOMs of these variables from a given Hamiltonian and Lindblad super operator, bypassing the usual interim step of the Fokker-Planck EOM for the quasi-probability density distribution function. Applying the so-called diffusion gauge (see Appendix \ref{sec:appA-III}) we obtain the following set of stochastic equations of motion:
\begin{gather}
	\frac{d \alpha(t)}{dt}  = - i \omega_c \alpha(t) - i f\sqrt{N} \rho_{eg}(t)
	- i \sqrt{\rho_{ee}(t)} F(t) -i \rho_{eg}(t) S(t), \nonumber \\
	\frac{d \alpha^\dagger(t)}{dt}  = ~~i \omega_c \alpha^\dagger(t) + i f\sqrt{N} \rho_{ge}(t)	
	+ i \sqrt{\rho_{ee}(t)} F^\dagger(t) +  i \rho_{ge} (t) S(t), \nonumber \\
	\frac{d \rho_{ee}(t)}{dt}  = - \gamma \rho_{ee}(t) + i \frac{f}{\sqrt{N}} \Bigl( \rho_{eg}(t) \alpha^\dagger(t) - \rho_{ge}(t) \alpha(t) \Bigr)
	- \frac{f}{\sqrt{N}} \rho_{ee}(t) S^*(t),
\label{eqn:final}\\
	\frac{d \rho_{eg}(t)}{dt}  = - \frac{\gamma}{2} \rho_{eg}(t) - i \omega_0 \rho_{eg}(t) + i \frac{f}{\sqrt{N}} \Bigl( 2 \rho_{ee}(t) - 1\Bigr) \alpha(t) 
	+ \frac{f}{\sqrt{N}} \sqrt{\rho_{ee}(t)} F^{\dagger *}(t) - \frac{f}{\sqrt{N}} \rho_{eg}(t) S^*(t), \nonumber\\
	\frac{d \rho_{ge}(t)}{dt} = - \frac{\gamma}{2} \rho_{ge}(t) + i \omega_0 \rho_{ge}(t) - i \frac{f}{\sqrt{N}} \Bigl( 2 \rho_{ee}(t) - 1\Bigr) \alpha^\dagger(t)  
	+ \frac{f}{\sqrt{N}} \sqrt{\rho_{ee} (t)} F^*(t) - \frac{f}{\sqrt{N}} \rho_{ge}(t) S^*(t), \nonumber
\end{gather}
\noindent where $F(t)$, $F^\dagger(t)$, and $S(t)$ are independent complex valued Gaussian white noise terms, satisfying the correlation properties:
\begin{equation}
	\langle F(t) F^* (t') \rangle = \delta(t- t'), \quad \langle F^\dagger (t) F^{\dagger*} (t') \rangle = \delta(t- t'), \quad \langle S (t) S^* (t') \rangle = \delta(t- t').
\label{eqn:gamfinal}
\end{equation}
Note, that due to the introduction of the stochastic terms the stochastic variables $\alpha$/$\alpha^{\dagger}$ and $\rho_{e,g}$/$\rho_{g,e}$ 
As highlighted by the derivation in Appendix \ref{sec:appA-I}, equations (\ref{eqn:quantum}) and (\ref{eqn:final}) have similar structure. Omitting the noise terms in (\ref{eqn:final}) leads to the Maxwell-Bloch equations, in equivalence to factorization of the second order correlations in eq. (\ref{eqn:quantum}). Hence, the noise terms in (\ref{eqn:final}) sample the contribution of the second-order correlators within the equations (\ref{eqn:quantum}), which are responsible for the quantum effects such as collapses and revivals of the population inversion in the TCM. In-depth discussion of the noise terms is given in Appendix \ref{sec:appE}.\\
Quantum mechanical averages are obtained by ensemble averages over stochastic realizations, for example
\begin{equation}
	\langle \hat{\sigma}_{ee} \rangle = \langle \rho_{ee}(t) \rangle = \frac{1}{n_\mathrm{traj}} \sum_{\mathrm{traj}} \rho_{ee}(t),
\end{equation}
\noindent where $n_\mathrm{traj}$ is the total amount of the sampled stochastic trajectories.

\section{Numerical parameter study and discussion}
\label{sec:3}

\subsection{Numerical simulation}
\label{sec:3-1}

In this section we discuss the numerical solution of (\ref{eqn:final}) and compare the results (where available) with the exact solution. The latter is obtained by means of: (i) numerical diagonalization of the Hamiltonian (\ref{H_TCM_proj}) in the case of the closed model ($\gamma = 0$); (ii) second quantization approach described in \cite{gegg_psiquasplibrary_2017} for the open system ($\gamma > 0$). For solving the SDEs we apply stochastic second order Runge-Kutta method \cite{milstein_stochastic_2004}. Statistical sampling is performed by averaging over $n_\mathrm{traj}=100,000$ trajectories.

We consider the particular case of the field in resonance with the atomic transition ($\omega_0 = \omega_c$).  We remove highly oscillating terms by means of the following transformation of variables:
\begin{gather}
	\alpha(t) \, \rightarrow \, \alpha(t) e^{-i \omega_0 t},  \quad \alpha^\dagger (t) \, \rightarrow \, \alpha^\dagger(t) e^{i \omega_0 t}, \nonumber\\
	\rho_{eg} (t) \, \rightarrow \, \rho_{eg} (t) e^{-i \omega_0 t}, \quad \rho_{ge} (t) \, \rightarrow \, \rho_{ge} (t) e^{i \omega_0 t}.
\end{gather}
\noindent Furthermore, we introduce the dimensionless time $\tau = f t$, thus making the obtained results independent of $f$ (assuming renormalization of the decay constant $\gamma \, \rightarrow \, \frac{\gamma}{f}$).
Initially the atomic and field subsystems are supposed to be uncorrelated. All atoms are supposed to start either from the ground or excited state. The initial condition of the field is chosen as a coherent state with average photon number $n_\mathrm{ph}$. In particular we consider two cases: $N = 10$ (small system) and $N = 100$ (large system). Their evolution is considered for the initial average photon numbers  $n_{\mathrm{ph}} = 0$ (vacuum), $n_{\mathrm{ph}} = 0.1 N$ ($n_{\mathrm{ph}} \ll N$), $n_{\mathrm{ph}} = N$ ($n_{\mathrm{ph}} \sim N$), and $n_{\mathrm{ph}} = 10 N$ ($n_{\mathrm{ph}} \gg N$).  
In the semi-classical limit of vanishing noise terms, the evolution of the atomic state population is governed by periodic Rabi oscillations. Inclusion of the noise terms, and thus quantum effects, changes the overall behavior of the solutions dramatically, leading to the well known phenomenon of the quantum collapse. Fig.\ref{fig:1_down} depicts three individual stochastic trajectories of the excited-state population alongside with the exact solution. The collapse of the population of the excited state is evident in the exact solution, less so in individual trajectories. For the collapse to be represented in the ensemble average over trajectories, destructive summation of oscillations of individual trajectories would be necessary. Notably, one of the displayed trajectories (dotted blue) rapidly reaches the numerical infinity. These so-called "runaway" trajectories are frequently observed in stochastic phase-space sampling approaches and lead to diverging ensemble averages. The time-points $t_S$ at which divergence is observed are classified as so-called movable singularities of the SDEs \cite{deuar_first-principles_2006}. In the vicinity of these points $t_S$ the individual solutions of (\ref{eqn:final}) have the form
\begin{equation}
	\rho_{ee} (t) = \frac{\text{const}}{(t - t_S)^2}, 
	\label{eqn:singularity}
\end{equation}
\noindent where $t_S$ depends on the particular realization of the stochastic trajectory. Singularities like these arise due to the general complex-valued noise terms multiplied with products of dynamical variables in eq. (\ref{eqn:gamfinal}). The divergences appear due to the breaking of hermicity of the stochastic dynamical variables (inherently linked to a doubling of the dimension of the Liouville space for the positive P distribution). This inherent issue of the stochastic approach, cannot be resolved by increasing the ensemble of trajectories, because its nature lies beyond the dispersive behavior of the stochastic trajectories and will be discussed in the next subsection. 

\begin{figure*}[t]
	\includegraphics[width=0.99\linewidth]{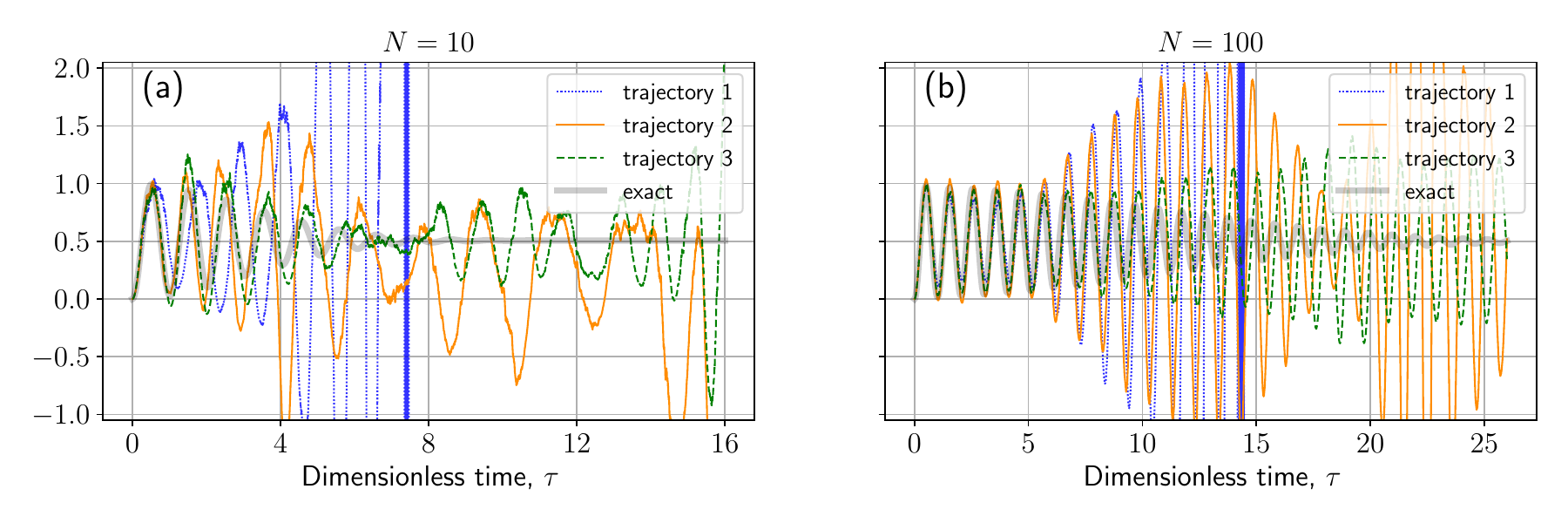}
	\caption{(Color online) Individual trajectories for the stochastic variable $\rho_{ee}$ and exact solution for the excited state probability of the closed TCM as a function of dimensionless time. Simulation parameters: $n_{ph} = 10N$, $\gamma/f =0$, atoms start from ground state; (a) $N = 10$ atoms; (b) $N = 100$ atoms.}
	\label{fig:1_down}
\end{figure*}

The situation is similar for the case of the closed TCM with $N = 10$ atoms starting from the ground state. In order to explore the behavior of the averages beyond the time point $t_S$, at which the first trajectory exhibits divergence, we manually remove the runaway trajectories for each time step and average the solutions over the remaining trajectories. The corresponding results are shown in the top and middle rows of Fig.\ref{fig:2}. The vertical dashed lines represent the time point for which $0.5\%$ of the initial trajectories have diverged. Before this critical time point, the averages from the stochastic sampling coincide with the exact results. Beyond this threshold value a large discrepancy between the averaged stochastic and the exact results is observed. We conclude that removal of a limited, small percentage of runaway trajectories can generate the exact solution for small propagation times. Generally, however, such a procedure leads to incorrect results and can therefore not be adopted for regularizing the divergences \cite{gilchrist_positive_1997}.

\begin{figure*}
	\includegraphics[width=0.99\linewidth]{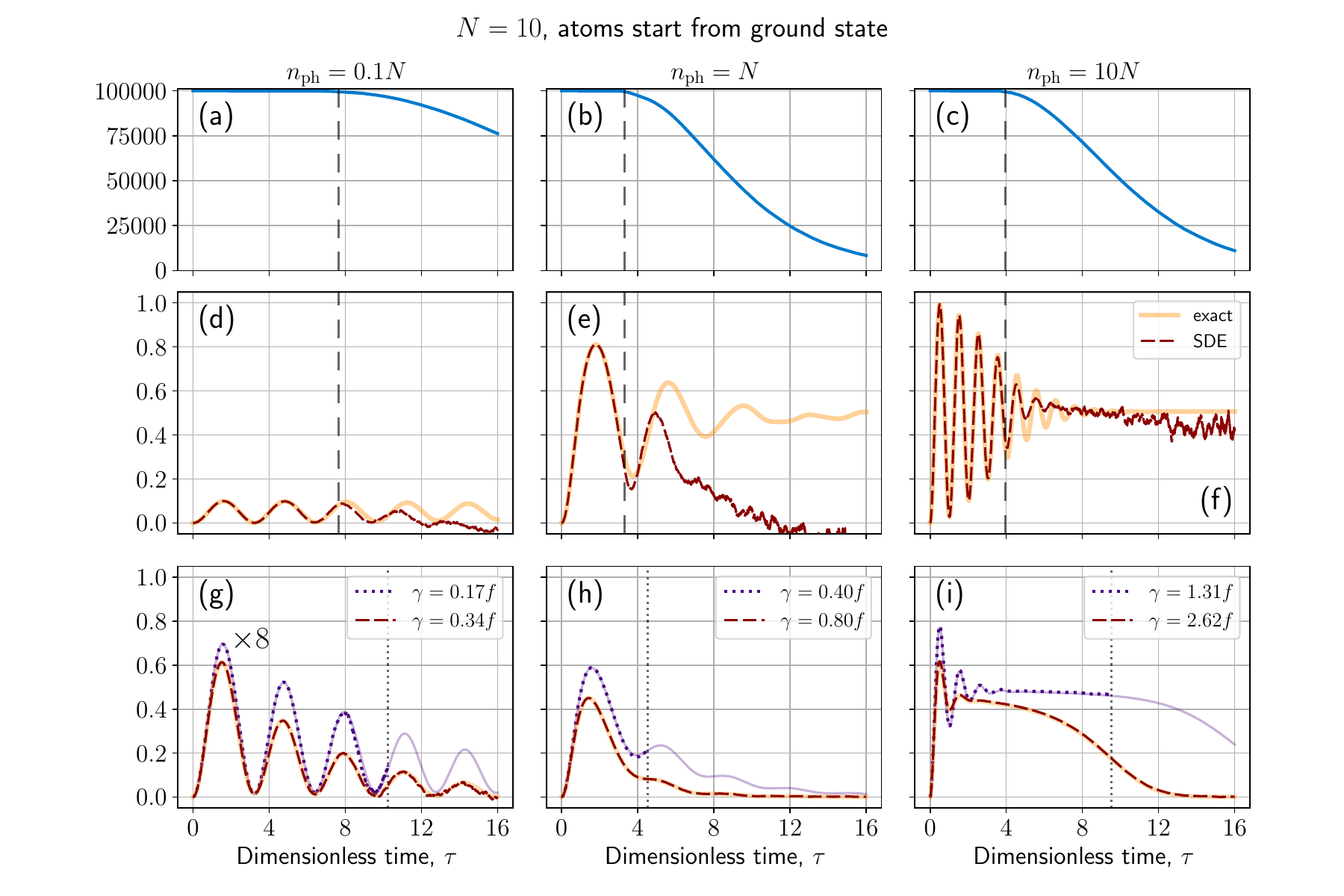}
	\caption{(Color online) (Top row) Amount of remaining trajectories and (Middle row) average of the stochastic variable $\rho_{ee}$ as a function of dimensionless time for the closed TCM ($\gamma = 0$). (Bottom row) Average of the stochastic variable $\rho_{ee}$ as a function of dimensionless time for the open TCM (values of $\gamma$ are given in the legend of each plot: bottom value corresponds to the minimal $\gamma$ leading to the converged results, top value is one half of the bottom value). Simulation parameters: $N=10$, all atoms start from the ground state and (a), (d), (g) $n_\text{ph} = 0.1N$; (b), (e), (h) $n_\text{ph} = N$; (c), (f), (i) $n_\text{ph} = 10N$. The vertical dashed line in (a)--(f) and dotted line in (g)--(i) show the time instant, at which the threshold of $0.5\%$ runaway trajectories has been reached. Solid lines correspond to the exact solution. Values in (g) are magnified x8 times.}
	\label{fig:2}
	\includegraphics[width=0.99\linewidth]{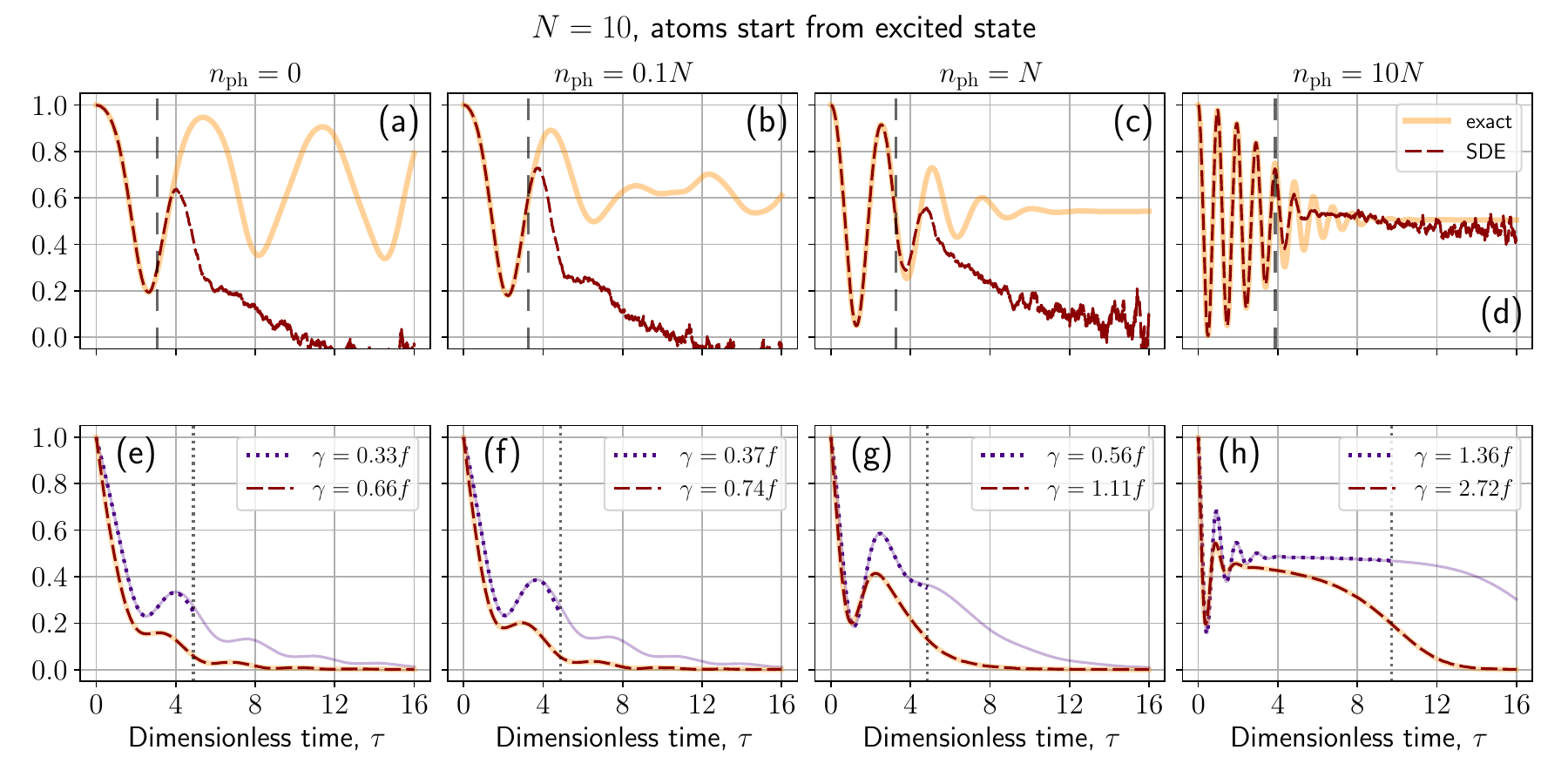}
	\caption{(Color online) Average of the stochastic variable $\rho_{ee}$ as a function of dimensionless time for the (Top row) closed TCM ($\gamma = 0$) and (Bottom row) open TCM (values of $\gamma$ are given in the legend of each plot: bottom value corresponds to the minimal $\gamma$ leading to the converged results, top value is one half of the bottom value). Simulation parameters: $N=10$, all atoms start from the excited state and (a), (g) $n_\text{ph} = 0$; (b), (h) $n_\text{ph} = 0.1N$; (c), (i) $n_\text{ph} = N$; (d), (j) $n_\text{ph} = 10N$. The vertical dashed in (a)--(d) and dotted in (e)--(h) lines show the time instant, at which the threshold of $0.5\%$ runaway trajectories has been reached. Solid lines correspond to the exact solution. }
	\label{fig:3}
\end{figure*}

\begin{figure*}
	\includegraphics[width=0.99\linewidth]{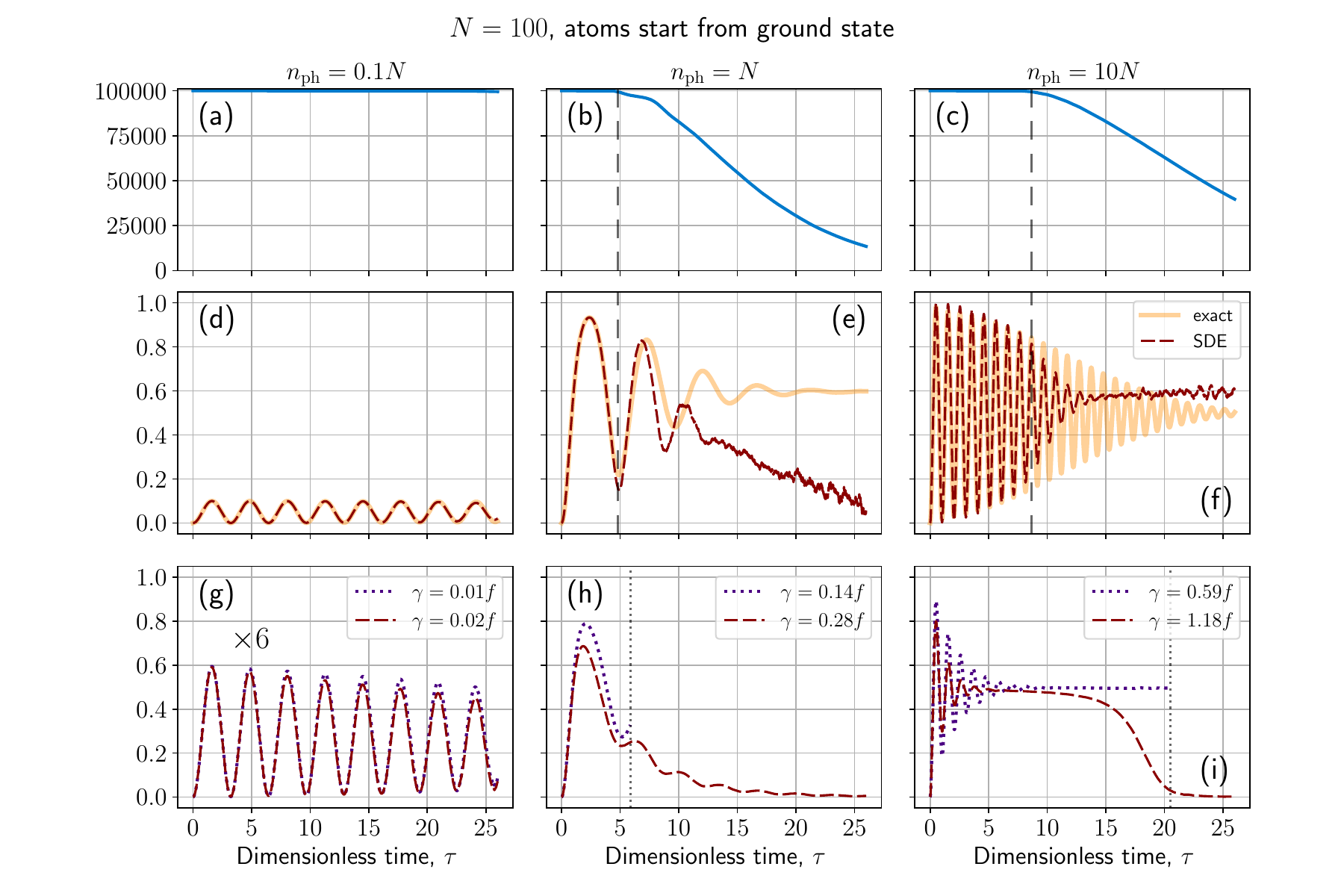}
	\caption{(Color online) (Top row) Amount of remaining trajectories and (Middle row) average of the stochastic variable $\rho_{ee}$ as a function of dimensionless time for the closed TCM ($\gamma = 0$). (Bottom row) Average of the stochastic variable $\rho_{ee}$ as a function of dimensionless time for the open TCM (values of $\gamma$ are given in the legend of each plot: bottom value corresponds to the minimal $\gamma$ leading to the converged results, top value is one half of the bottom value). Simulation parameters: $N=100$, all atoms start from the ground state and (a), (d), (g) $n_\text{ph} = 0.1N$; (b), (e), (h) $n_\text{ph} = N$; (c), (f), (i) $n_\text{ph} = 10N$. The vertical dashed in (a)--(f) and dotted in (h)--(i) lines show the time instant, at which the threshold of $0.5\%$ runaway trajectories has been reached. Solid lines correspond to the exact solution. Values in (g) are magnified x6 times and included as an example of open TCM. }
	\label{fig:6}
	\includegraphics[width=0.99\linewidth]{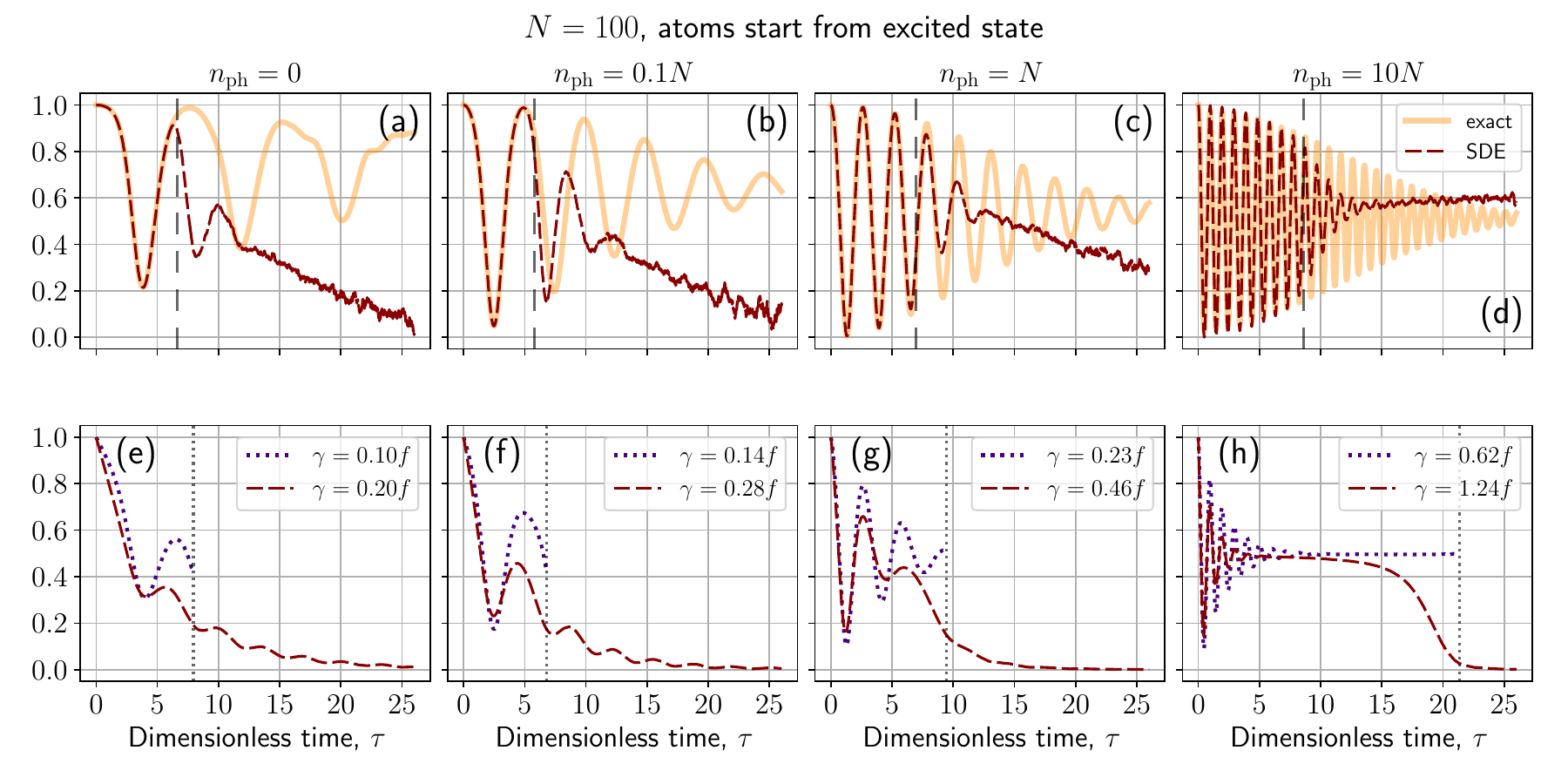}
	\caption{(Color online) Average of the stochastic variable $\rho_{ee}$ as a function of dimensionless time for the (Top row) closed TCM ($\gamma = 0$) and (Bottom row) open TCM (values of $\gamma$ are given in the legend of each plot: bottom value corresponds to the minimal $\gamma$ leading to the converged results, top value is one half of the bottom value). Simulation parameters: $N=100$, all atoms start from the excited state and (a), (g) $n_\text{ph} = 0$; (b), (h) $n_\text{ph} = 0.1N$; (c), (i) $n_\text{ph} = N$; (d), (j) $n_\text{ph} = 10N$. The vertical dashed in (a)--(d) and dotted in (e)--(h) lines show the time instant, at which the threshold of $0.5\%$ runaway trajectories has been reached. Solid lines correspond to the exact solution. }
	\label{fig:7}
\end{figure*}

It has been observed, that convergence of the stochastic ensemble averages can be reached when the volume of the phase space decreases at a rate exceeding the rate of approaching the numerical infinities by the runaway trajectories \cite{mandt_stochastic_2015,naether_stationary_2015,deuar_fully_2021}. A decrease of the effective phase-space volume can be established by introduction of damping. We thus introduce damping and establish the minimal value of the normalized dissipation parameter $\gamma/f$, which regularizes the behavior of the trajectories. The corresponding results for $N=10$ are shown by the dashed lines in the bottom row of Fig.\ref{fig:2}. Furthermore, we show results for $0.5\gamma/f$ (shown by the dotted lines) up to propagation times for which $0.5\%$ runaway trajectories are reached. We observe that the value of $\gamma/f$ strongly depends on the amount of initial energy. An increase of the number of the field quanta results in an increase of the total number of excitations in the system, which takes longer times to being damped by the dissipation channel. The singularity is, hence, often reached before a noticeable decrease of the phase space volume. The required $\gamma/f$ value for effective regularization is in most cases quite large and drastically changes the evolution of the closed system. Fig.\ref{fig:3} shows simulation results for the same parameters as in Fig.\ref{fig:2} but with all the atoms starting in the excited state. As expected, the averages of the SDE solutions diverge even more rapidly compared to the atomic system starting from the ground state. In conclusion, introduction of a damping term for regularization of runaway trajectories is generally inapplicable.

A parameter study for large systems ($N=100$) is shown in Fig.\ref{fig:6}--\ref{fig:7}.  Keeping the total number of excitations relatively low ($\overline{n}_{ph} \lesssim N$), the range of validity of the stochastic method is extended for the closed TCM.  As seen in Fig.\ref{fig:6}d, we manage to obtain convergence within the considered time interval for the case $n_\mathrm{ph} = 0.1N$ and atoms starting from the ground state. The noise terms scale with $1/\sqrt{N}$ and keep the evolution closer to the semi-classical limit for large systems. Furthermore, the values of the regularizing $\gamma/f$ parameter decrease with increasing $N$, as seen in Fig.\ref{fig:2}--\ref{fig:3}.  Switching to an initially completely inverted system introduces more excitations and requires higher damping rates counteracting the stochastic contribution.

\subsection{The possible nature of the divergence}
\label{sec:3-2}

Omitting the noise terms, the equations of motion (\ref{eqn:final}) are non-linear. Their solution is, however, regular and non-divergent, assuming physically meaningful initial conditions and the variables pertaining to Hermitian conjugate operators being conjugate transpose to each other. For large system size $N \gg 1$, the contribution of the stochastic terms (scaling with $1/N$) can be considered small compared to the deterministic part. A reasonable question arising is therefore, why inclusion of the noise terms leads to such drastic changes of the obtained solution. In order to find an explanation, we consider the solution of the deterministic part of eqs.\ (\ref{eqn:final}) for the closed TCM ($\gamma$=0).

First, let us stress that assuming the semi-classical initial conditions $\alpha^\dagger (0) = \alpha^* (0)$, $\rho_{ge} (0) = \rho_{eg}^* (0)$, and $\rho_{ee} (0) \in [0,1]$ leads to the following identities: 
\begin{equation}
	\alpha^\dagger(t) = \alpha^*(t); \quad \rho_{ij}(t) = \rho^*_{ji}(t).
\end{equation}
In the next step, we derive an EOM for the pertaining one d.o.f. To that end, we differentiate the equation for $\rho_{ee}(t)$ to obtain
\begin{equation}
	\frac{d^2 \rho_{ee} (t)}{dt^2} = - 2 f^2 \Bigl[ \bigl( 2 \rho_{ee} (t)  - 1 \bigr)\frac{\alpha^\dagger(t) \alpha(t)}{N}  + \rho_{eg}(t) \rho_{ge}(t) \Bigr].
	\label{eqn:pee2}
\end{equation}
\noindent Taking into account the semi-classical conservation for the total energy $E$ of the TCM (its value we normalize over $N$) and the unit length of the Bloch vector, we obtain:
\begin{gather}
	\frac{ \bigl( 2 \rho_{ee} (t) - 1 \bigr)}{2} + \frac{\alpha^\dagger(t) \alpha(t)}{N} = E, \nonumber\\
	\bigl( 2 \rho_{ee}(t) - 1 \bigr)^2 + 4 \rho_{eg}(t) \rho_{ge}(t) = 1.
\label{eqn:e-bl}
\end{gather}
As a result, one can derive the following second order ODE
\begin{equation}
	\frac{1}{f^2} \frac{d^2 w(t)}{dt^2} \equiv \ddot{w} = - 4 E w(\tau)  + 3 w(\tau)^2 - 1,
\label{eqn:anh_osc}
\end{equation}
\noindent where $w(\tau) = \bigl(2 \rho_{ee}(\tau) - 1\bigr)$ is the population inversion, $\ddot{w}(\tau) \equiv \frac{d^2 w(\tau)}{d \tau^2}$, and $\tau = f t$ is the dimensionless time. Integrating  this equation once in time, we obtain:
\begin{equation}
	\frac{\dot{w}(\tau)^2}{2} = - 2 E w(\tau)^2 + w(\tau)^3 - w(\tau) + h_w, \quad\quad  h_w = \text{const}.
\label{eqn:class_ham}
\end{equation}
A closer look at equation (\ref{eqn:class_ham}) allows one to interpret the constant of integration $h_w$ as an effective Hamiltonian of the following classical conservative system moving in a 1-D anharmonic potential:
\begin{equation}
	h_w (w, \dot{w}) = \frac{\dot{w}^2}{2} + U(w), \quad\quad U(w) = 2 E w^2 - w^3 + w,
\label{eqn:eff_potential}
\end{equation}
\noindent where the variable $w(\tau)$ plays the role of an effective generalized coordinate. The harmonic part of the potential has the eigenfrequency $\Omega = \sqrt{4 E}$, and for initial pure quantum states of the atomic subsystem $w(0) = \pm 1$ the quantity $h_w=2E$.  Fig.\ref{fig:9} shows the shape of potential (\ref{eqn:eff_potential}), which contains both the regions of   bound and unbound motion. Being in the semi-classical regime of evolution restricts the possible value of the effective energy $h_w$, leading to oscillations of the population inversion between the values $\pm 1$.

\begin{figure*}[!t]
	\includegraphics[width=1.0\linewidth]{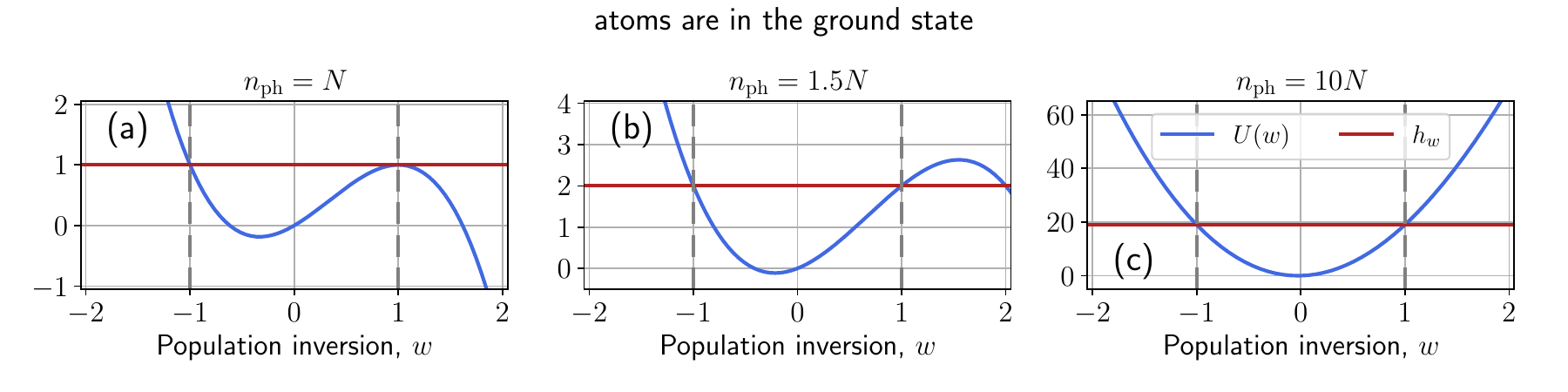}
	\caption{(Color online) Effective potential and effective Hamiltonian (\ref{eqn:eff_potential})  as a function of the population inversion $w$.}
	\label{fig:9}
\end{figure*}

However, including the noise terms in the equations of motion leads to the violation of the energy conservation law for individual trajectories and in some cases eventually resulting in an increased value above  the potential barrier. As a result, the system transits to  unbounded values of $w$. Moreover, when $|w| \gg 1$ the cubic term in (\ref{eqn:class_ham}) dominates, so that we can approximately deduce:
\begin{equation}
 	\dot{w}(\tau)^2 \sim w(\tau)^3 \quad \Rightarrow \quad \dot{w}(\tau) \sim w(\tau)^{\frac{3}{2}} \quad \Rightarrow \quad \frac{1}{\sqrt{w(\tau)}} \sim (\tau - \tau_S) \,\,\, \text{or} 
 		\,\,\, w(\tau) \sim \frac{1}{(\tau - \tau_S)^2},
 \label{eqn:osc_singlular}
 \end{equation} 
\noindent which leads to the singularity (\ref{eqn:singularity})  mentioned in the previous subsection.

We can also provide a qualitative explanation of the rapid behavior of such divergence by rearranging the terms in equation (\ref{eqn:anh_osc}) and approximating them by:
\begin{equation}
	\ddot{w}(\tau) =  -\biggl(4E - 3w(\tau) \biggr)w(\tau) - 1 \approx - (\Omega^2 + z_0) w(\tau).
\end{equation} 
\noindent The noise terms being complex implies that the value $z_0$ also becomes complex. Its real part affects the frequency of the effective oscillator whereas its imaginary part leads to the exponential growth of the amplitude. When the imaginary part becomes noticeable, the system quickly approaches the behavior given by equation (\ref{eqn:osc_singlular}).

\subsection{drift-gauge transformations}
\label{sec:3-3}

SDEs possess a general property, allowing to change the deterministic ("drift") part of the equations under the condition of keeping the expectation values unaffected by the transformation. This technique is called drift-gauge transformation \cite{deuar_thesis_2005} and in some cases can help to regularize divergent behaviour of the SDEs. The general scheme for its implementation requires the introduction of an additional stochastic variable $\Omega(t)= e^{C_0 (t)}$ that serves as a weight function to maintain the correct expectation values. An arbitrary dynamical variable $\mathcal{O}(t)$ changes to $\widetilde{O}(t)$ upon the transformation by the following equations:
\begin{gather}
	\langle \mathcal{O}(t) \rangle \, \longrightarrow  \,\langle \mathcal{\widetilde{O}}(t) \Omega (t) \rangle,
\label{d_gauge_av} \\
	\left( \frac{d \mathcal{O}}{d t} \right)_\text{det} \, \longrightarrow  \, \left( \frac{d \mathcal{\widetilde{O}}}{d t} \right)_\text{det} =
		\left( \frac{d \mathcal{O}}{d t} \right)_\text{det} + \Delta \mathcal{O}(t), \nonumber
\end{gather}
\noindent where $\Delta \mathcal{O}(t)$ is a deterministic shift of the stochastic equation for $\mathcal{O}(t)$. The stochastic exponent  $C_0(t)$ of the weight function ist determined by the following equation
\begin{equation}
\frac{d C_0 (t) }{d t} = \left( \frac{d C_0}{d t} \right)_\text{det} + \left( \frac{d C_0}{d t} \right)_\text{stoch}\;, 
\end{equation}
which implicitly depends on $\Delta \mathcal{O}(t)$ (for more details see Appendix \ref{sec:appD}). 

As previously discussed, the contribution of runaway trajectories to the averages is essential and manual removal of runaway trajectories from the sampling procedure leads to the incorrect results \cite{gilchrist_positive_1997}. Taking into account the findings of  Sec.\ref{sec:3-2}, we explore a possible drift-gauge transformation that effectively transforms the potential (\ref{eqn:eff_potential}) in such a way that motion of the ''generalized coordinate'' $w(t)$ is bounded for any initial value of the parameter $h_w$. In particular, this is achieved by artificially introducing an additional term $\sim w^4$ within the effective potential (\ref{eqn:eff_potential}) properly scaled to outcompete the negative cubic term. To the end we modify the equations for the variables $\alpha(t)$ and $\alpha^\dagger (t)$ as follows (for details see Appendix \ref{sec:appD}):
\begin{gather}
	\Delta \alpha(t) = \varkappa \cdot i f\sqrt{N} \rho_{eg}(t) \re{[\rho_{ee}(t)]}, \nonumber\\
	\Delta \alpha^\dagger(t) = - \varkappa \cdot i f\sqrt{N} \rho_{ge}(t) \re{[\rho_{ee}(t)]},
\label{eqn:a-shifts}
\end{gather}
\noindent where $\varkappa$ can be any real-valued, positive function of time and other variables (below we call it the switch function). Furthermore, the transformation only depends on the real part of the stochastic variable $\rho_{ee} (t)$, thereby modyfying the real-valued potential (\ref{eqn:eff_potential}) and bounding the motion of the real part of $\rho_{ee}(t)$ in the semi-classical equations. This results in the following modification of the effective potential (\ref{eqn:eff_potential}) (note that in the semi-classical case $\rho_{ee} (t) \equiv \re{[\rho_{ee}(t)]}$):
\begin{equation}
	U(w) = \Bigl(2 E - \frac{\varkappa}{8} \Bigr)w^2 - \Bigl(1 - \frac{\varkappa}{8} \Bigr) w^3 + \Bigl( 1 + \frac{\varkappa}{8}\Bigr) w + \frac{\varkappa}{32} w^4.
	\label{eqn:eff-pot-mod}
\end{equation}
The required weight function $C_0(t)$ depends on the correlators between $\alpha(t)$, $\alpha^\dagger(t)$ and $C_0(t)$. Remarkably, the transformation (\ref{eqn:a-shifts}) does not result in additional noise terms or changes thereof in the EOM for $\alpha(t)$, $\alpha^\dagger(t)$. As a result, we derive the following EOMs.
\begin{gather}
	\frac{d \widetilde{\alpha}(t)}{dt}  = - i \omega_c \alpha(t) - i f\sqrt{N} \rho_{eg}(t) \Bigl( 1 - \varkappa \re{[\rho_{ee}(t)]} \Bigr) 
	- i \sqrt{\rho_{ee}(t)} F(t) -i \rho_{eg}(t) S(t), \nonumber \\
	\frac{d \widetilde{\alpha}^\dagger(t)}{dt}  = ~~i \omega_c \alpha^\dagger(t) + i f\sqrt{N} \rho_{ge}(t) \Bigl( 1 - \varkappa \re{[\rho_{ee}(t)]} \Bigr) 	
	+ i \sqrt{\rho_{ee}(t)} F^\dagger(t) +  i \rho_{ge} (t) S(t), 
	\label{eqn:g-transform}\\
	\frac{d C_0(t)}{dt} =  \varkappa \cdot \sqrt{N} \re{[\rho_{ee}(t)]} S^*(t). \nonumber
\end{gather}
\textbf{}

Fig.\ref{fig:10}a shows the results of the numerical solution of (\ref{eqn:final}) with transformation (\ref{eqn:g-transform}) for the closed TCM ($\gamma =0$) with $N=10$ atoms. We compare the averages of the SDEs to the exact solution for $\rho_{ee}(t)$. Compared to Fig.\ref{fig:2}f,  runaway trajectories partly disappear and move to the later simulation times. This is shown by the vertical dotted line, which corresponds to the moment of time when more than 0.5\% of the trajectories run out of the pre-defined bounds. The fact that runaway trajectories are not completely removed is connected to the complex noise terms, which introduce a finite imaginary part of $\rho_{ee}(t)$. The evolution of the imaginary part of $\rho_{ee}(t)$ is generally not bounded by the potential (\ref{eqn:eff-pot-mod}). Furthermore, this imaginary part reenters the EOMs of $\rho_{ee}(t)$ and other variables as a multiplicative factor of the complex-valued noise terms, eventually significantly augmenting the real part of $\rho_{ee}(t)$ and leading to numerical infinities. This corresponds to the rapid increase of $h_w$ along the potential walls of (\ref{eqn:eff-pot-mod}). 

The weight function itself, however introduces divergences. This is highlighted by the vertical dashed line in Fig.\ref{fig:10}a, which corresponds to the moment of time when more than 0.5\% weighted trajectories ran out of the simulation bounds. Remarkably, its position is almost congruent with the position of the vertical line in Fig.\ref{fig:2}f. The reason behind such a behaviour is that the evolution of the exponent $C_0(t)$ of the weight function is determined by an independent multiplicative Wiener process, whereas the variance of $C_0(t)$ even in the simplest case of additive noises is $\sim \sqrt{t}$. This implies a variance of the weight function to be at least $\sim e^{\sqrt{t}}$.

To challenge the exponential growth of both the weight function and its variance, we tried to keep the right-hand side of the SDE for $C_0(t)$ in (\ref{eqn:g-transform}) turned off (i.e., equal to zero or nearly zero) as long as possible and turned it on adiabatically only once the real part of the variable $\rho_{ee}(t)$ reaches a certain threshold. Mathematically we tried to realize that by introducing the explicit form of the switch function $\varkappa$ in the way as follows:
\begin{equation}
	\varkappa (x) = 1 + \frac{1}{2} \Bigl(\tanh{\bigl[k(x_1 - x)\bigr]} + \tanh{\bigl[k(x - x_2)\bigr]} \Bigr),
\label{eqn:switch}
\end{equation}
\noindent where $x = \re{[\rho_{ee}(t)]}$; $k,x_1,x_2$ are the parameters of the switch function. In order to affect only those trajectories, in which values for $\re{[\rho_{ee}(t)]}$ tend to leave the region $[0,1]$, we chose these parameters as: $x_1 = -1$, $x_2 = 2$, $k = 1$.
 
Fig.\ref{fig:10}c shows the correlation in evolution of $\rho_{ee}(t)$, $C_0(t)$, and $\varkappa(t)$ for a single, unbounded trajectory of a closed TCM with parameters from Fig.\ref{fig:10}a. Unfortunately, in the case of a closed TCM $\re{[\rho_{ee}(t)]}$ frequently reaches the threshold values, leading to the noticeable turned on time of the switch function. As a result, significant runaway of the  weight function manages to take place.

 Fig.\ref{fig:10}b shows the evolution of the open TCM  for the parameters of Fig.\ref{fig:10}a and $\gamma /f = 2.6$. Furthermore Fig.\ref{fig:10}b highlights the importance of the weight function in the averaging procedure (\ref{d_gauge_av}) by showing properly weighted  ensemble averages ("SDE") and averages calculated without the weight function ("SDE (n/w)"). Notably, the non-weighted averages cannot produce the exact results. In addition, Fig.\ref{fig:10}d shows the correlation of the evolution of $\rho_{ee}(t)$ and $C_0(t)$ with the switch function $\varkappa (t)$ for a single trajectory in the case with sufficient damping ($\gamma /f = 2.6$).

\begin{figure*}[!t]
	\includegraphics[width=0.94\linewidth]{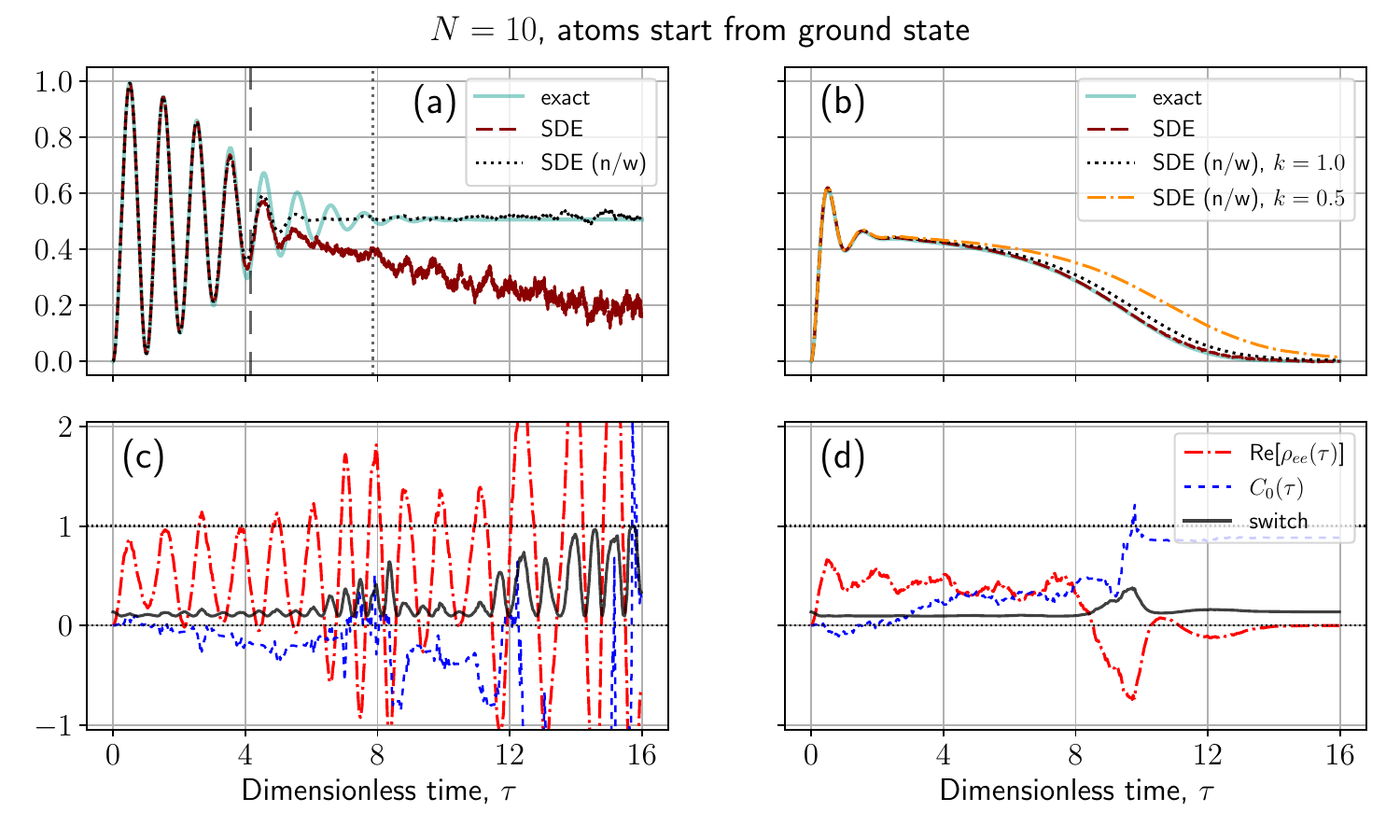}
	\caption{(Color online) (Top row) The weighted and non-weighted averages of the stochastic variable $\rho_{ee}$ following the eqns. (\ref{eqn:final}), (\ref{eqn:g-transform}) and (Bottom row) real value of $\rho_{ee}$, $C_0$ and switch function (\ref{eqn:switch}) of a single trajectory as a function of  the dimensionless time. Simulation parameters: $N=10$, $n_{\text{ph}} = 10N$ and (a), (c) $\gamma/f=0$; (b), (d) $\gamma=2.6f$. Parameters of switch function (\ref{eqn:switch}): $x_1 = -1$, $x_2 = 2$, $k = 1$.}
	\label{fig:10}
\end{figure*}

\section{Conclusions}
\label{sec:4}

In our work, we analyzed the critical points of the stochastic trajectory approach applied to the quantum dynamics of both the closed and the open Tavis-Cummings model. 
We showed that for the closed and weakly dissipative, open TCM the resulting stochastic trajectories exhibit intrinsic divergent behavior in the form of runaway trajectories. Removing such trajectories from the calculated averages beyond the threshold of about $0.5\%$ leads to incorrect results. We showed that the minimum value of regulating dissipation parameter leading to non-diverging results strongly depends on the number of initial excitations in the system. Furthermore, we analyzed the mathematical reason for the presence of runaway trajectories. The proposed drift-gauge transformation of the SDEs proofed ineffective to regulate the divergent behavior of the SDEs.

Consequently, we conclude that the formulation of the stochastic trajectory approach is best suited for strongly dissipative TCM systems with a large number of emitters and low initial excitation values. Only in this limiting case the exact quantum dynamics of the system can be reproduced by the stochastic approach. Overall, we conclude that the stochastic method works well for systems approaching a semi-classical limit, but fails to describe long-time evolution of quantum effects.

\section{Acknowledgements}

Authors would like to acknowledge Michael Gegg for useful hints on running the PsiQuaSP code. This research was supported in part through the Maxwell computational resources operated at Deutsches Elektronen-Synchrotron DESY, Hamburg, Germany. Furthermore we want to acknowledge financial support of this work by the Grant-No. HIDSS-0002 DASHH (Data Science in Hamburg-Helmholtz Graduate School for the Structure of Matter).

\end{large}

\bibliography{bibljtcm}

\appendix

\section{Derivation of stochastic equations of motion}
\label{sec:appA}

In order to derive the SDEs governing the evolution of TCM, let us split each equation into the two parts -- deterministic (''det'') and stochastic (''stoch''):
\begin{equation}
	\frac{d \mathcal{O}(t)}{dt} = \left( \frac{d \mathcal{O}}{dt} \right)_\text{det} + \left( \frac{d \mathcal{O}}{dt} \right)_\text{stoch}, 
\label{eqnA:sde}
\end{equation}
\noindent where $\mathcal{O}(t)$ is an arbitrary stochastic variable. Deterministic part describes the semi-classical evolution of the system, whereas the stochastic part contains the noise terms with definite correlation properties and is responsible for the quantum effects. In order to derive (\ref{eqnA:sde}) explicitly, one may follow one of the two strategies described in the next subsections.

\subsection{Method I: Ansatz with stochastic sampling}
\label{sec:appA-I}

Before considering the general formalism of the positive P-representation and additional ways of handling the explicit form of the equations of motion, we would like to propose a simpler approach resulting in the same explicit form of SDEs \cite{chuchurka_ansatz}. Its main idea is inspired by the properties of Ito's calculus \cite{ito_stochastic_1944} and consists in using the following ansatz for the total density operator, which factorizes all degrees of freedom of the TCM:
\begin{gather}
	\hat{\rho}_{\text{ans}} (t) = \hat{\rho}_{A} (t) \cdot \hat{\rho}_{F} (t),
	\label{eqnA1:rho-tot} \\
	\hat{\rho}_{A} (t) = \prod_{\mu = 1}^{N} \hat{\rho}_{\mu} (t), \quad \hat{\rho}_{F} (t) = 
		\hat{\Lambda}\left(\alpha (t), \alpha^\dagger (t)\right) = e^{-\alpha^\dagger(t) \alpha(t)} e^{\alpha(t) \hat{a}^\dagger} \ket{0}\bra{0} e^{\alpha^\dagger(t) \hat{a}},
	\label{eqnA1:rho-fac}
\end{gather}
\noindent where $\hat{\rho}_{\mu} (t)$  is the one-particle atomic density operator, which can be expanded in terms of the $\hat{\sigma}$-projectors as
\begin{equation}
	\hat{\rho}_{\mu} (t) = \sum_{pq} \rho^{(\mu)}_{pq}(t) \hat{\sigma}^{(\mu)}_{pq}.
	\label{eqnA1:rho-at}
\end{equation}

One can easily prove the following identities:
\begin{gather}
	\rho^{(\mu)}_{pq} (t) = \Tr{\Bigl(\hat{\sigma}^{(\mu)}_{qp} \hat{\rho}_{\text{ans}} (t) \Bigr)}  =  \Tr_{\mu} {\Bigl(\hat{\sigma}^{(\mu)}_{qp} \hat{\rho}_{\mu} (t) \Bigr)},
	\qquad p,q \in \{e,g\};
\label{eqnA1:rhomu}\\
	\alpha (t) = \Tr{\Bigl(\hat{a} \hat{\rho}_{\text{ans}} (t) \Bigr)}  = \Tr_F{\Bigl(\hat{a} \hat{\rho}_F (t) \Bigr)}, \quad
		\alpha^\dagger (t) = \Tr{\left(\hat{a}^\dagger \hat{\rho}_{\text{ans}} (t) \right)}  = \Tr_F{\left(\hat{a}^\dagger \hat{\rho}_F (t) \right)}.
\label{eqnA1:aa}
\end{gather}
\noindent (Let us stress that the dagger symbol in $\alpha^\dagger (t)$ does not assume any mathematical operation and is used as a convenient notation).

The ansatz (\ref{eqnA1:rho-tot})--(\ref{eqnA1:rho-at}) also factorizes the second-order correlators:
\begin{gather}
	\Tr \Bigl( \hat{\sigma}^{(\mu)}_{qp} \hat{\sigma}^{(\nu)}_{rs} \hat{\rho}_{\text{ans}} (t)   \Bigr) = \rho^{(\mu)}_{pq} (t) \rho^{(\nu)}_{sr} (t), \nonumber \\
	\Tr \Bigl( \hat{a} \hat{\sigma}^{(\mu)}_{qp} \hat{\rho}_{\text{ans}} (t)   \Bigr) = \alpha (t) \rho^{(\mu)}_{pq} (t), \quad
		\Tr \Bigl( \hat{a}^\dagger \hat{\sigma}^{(\mu)}_{qp} \hat{\rho}_{\text{ans}} (t)   \Bigr) = \alpha^\dagger (t) \rho^{(\mu)}_{pq} (t).
\label{eqnA1:double}
\end{gather}

Let us differentiate (\ref{eqnA1:rhomu})--(\ref{eqnA1:aa}) over time. As a result, we obtain:
\begin{gather}
	\frac{d \rho^{(\mu)}_{pq} (t) }{dt} = \Tr{\Bigl(\hat{\sigma}^{(\mu)}_{qp} \dot{\hat{\rho}}_{\text{ans}} (t) \Bigr)} = \Tr{\Bigl(\hat{\sigma}^{(\mu)}_{qp} \mathcal{L} [\hat{\rho}_{\text{ans}} (t)]  \Bigr)}, \nonumber \\
	\frac{d \alpha (t)}{dt} = \Tr{\Bigl(\hat{a} \mathcal{L} [\hat{\rho}_{\text{ans}} (t)] \Bigr)}, \quad
		\frac{d \alpha^\dagger (t)}{dt} = \Tr{\Bigl(\hat{a}^\dagger \mathcal{L} [\hat{\rho}_{\text{ans}} (t)] \Bigr)}.	
\label{eqnA1:derivTr}
\end{gather}

Accounting for (\ref{eqnA1:rhomu})--(\ref{eqnA1:double}), one can derive (\ref{eqnA1:derivTr}) explicitly:
\begin{align}
	\left( \frac{d \alpha}{dt} \right)_\text{det} &= - i \omega_c \alpha(t) - i \frac{f}{\sqrt{N}} \sum_\nu \rho_{eg}^{(\nu)}(t), \nonumber \\
	\left( \frac{d \alpha^\dagger}{dt} \right)_\text{det} &= ~~i \omega_c \alpha^\dagger(t) + i \frac{f}{\sqrt{N}} \sum_\nu \rho_{ge}^{(\nu)}(t), \nonumber \\
	\left( \frac{d \rho_{ee}^{(\mu)}}{dt} \right)_\text{det} &= ~~i \frac{f}{\sqrt{N}} \left( \rho_{eg}^{(\mu)}(t) \alpha^\dagger(t) - \rho_{ge}^{(\mu)}(t) \alpha(t) \right)
		- \gamma \rho_{ee}^{(\mu)}(t),
\label{eqnA1:classic}\\
	\left( \frac{d \rho_{gg}^{(\mu)}}{dt} \right)_\text{det} &= - i \frac{f}{\sqrt{N}} \left( \rho_{eg}^{(\mu)}(t) \alpha^\dagger(t) - \rho_{ge}^{(\mu)}(t) \alpha(t) \right)
		+ \gamma \rho_{ee}^{(\mu)}(t), \nonumber\\
	\left( \frac{d \rho_{eg}^{(\mu)}}{dt} \right)_\text{det} &= - i \omega_0 \rho_{eg}^{(\mu)}(t) + i \frac{f}{\sqrt{N}} \left( \rho_{ee}^{(\mu)}(t) - \rho_{gg}^{(\mu)}(t)\right) \alpha(t) - \frac{\gamma}{2} \rho_{eg}^{(\mu)}(t), \nonumber\\
	\left( \frac{d \rho_{ge}^{(\mu)}}{dt} \right)_\text{det} &= ~~i \omega_0 \rho_{ge}^{(\mu)}(t) - i \frac{f}{\sqrt{N}} \left( \rho_{ee}^{(\mu)}(t) - \rho_{gg}^{(\mu)}(t)\right) \alpha^\dagger(t) - \frac{\gamma}{2} \rho_{ge}^{(\mu)}(t). \nonumber
\end{align}

Equations (\ref{eqnA1:classic}) describe evolution of the TCM with strong classical behavior (the latter is underlined by putting the extra ''det'' notation, which totally corresponds to the spirit of (\ref{eqnA:sde})).

Now let us derive the terms, which were neglected in operator equation (\ref{eq:masterdiss}) when deriving (\ref{eqnA1:classic}) and analyze their structure. Taking the time derivative of (\ref{eqnA1:rho-tot}), we obtain:
\begin{gather}
	\frac{d \hat{\rho}_{\text{ans}} (t)}{dt} = \frac{d \hat{\rho}_A(t) }{dt} \hat{\rho}_F(t) + \hat{\rho}_A(t) \frac{d \hat{\rho}_F(t) }{dt}, \nonumber \\
	\frac{d \hat{\rho}_A(t) }{dt} = \sum_{\mu = 1}^{N} \left( \prod_{\nu \neq \mu} \hat{\rho}_\nu(t)  \right) \sum_{pq} \dot{\rho}^{(\mu)}_{pq}(t) \hat{\sigma}^{(\mu)}_{pq},
	\quad \frac{d \hat{\rho}_F(t) }{dt} = \left( \dot{\alpha}(t) \frac{\partial}{\partial \alpha} + \dot{\alpha^\dagger}(t) \frac{\partial}{\partial \alpha^\dagger} \right)
		\hat{\Lambda} (\alpha(t), \alpha^\dagger(t)),
\label{eqnA1:genderiv}
\end{gather}
\noindent where we used the following identities \cite{drummond_generalised_1980}:
\begin{align}
	& \hat{a} \hat{\Lambda}\left(\alpha, \alpha^\dagger \right) = \alpha \hat{\Lambda}\left(\alpha, \alpha^\dagger \right), \nonumber\\
	& \hat{a}^\dagger \hat{\Lambda}\left(\alpha, \alpha^\dagger \right) = \left( \alpha^\dagger + \frac{\partial}{\partial \alpha} \right) \hat{\Lambda}\left(\alpha, \alpha^\dagger \right), \nonumber\\
	& \hat{\Lambda}\left(\alpha, \alpha^\dagger \right) \hat{a}^\dagger = \hat{\Lambda}\left(\alpha, \alpha^\dagger \right) \alpha^\dagger,\\
	& \hat{\Lambda}\left(\alpha, \alpha^\dagger \right) \hat{a} = \left( \frac{\partial}{\partial \alpha^\dagger} + \alpha \right) \hat{\Lambda}\left(\alpha, \alpha^\dagger \right). \nonumber
\end{align}

Substituting  equations (\ref{eqnA1:classic}) into (\ref{eqnA1:genderiv}), we derive:
\begin{eqnarray}
	\mathcal{L} [\hat{\rho}_{\text{ans}} (t)] - \frac{d \hat{\rho}_{\text{ans}}(t) }{dt} =
		\frac{i f}{\sqrt{N}} \sum_{\mu = 1}^{N} \left( \prod_{\nu \neq \mu} \hat{\rho}_\nu(t)  \right) \sum_{pq} \rho^{(\mu)}_{pq} (t)
			\left[ \rho^{(\mu)}_{eg}(t) \frac{\partial}{\partial \alpha} - \rho^{(\mu)}_{ge}(t) \frac{\partial}{\partial \alpha^\dagger} \right]
			\hat{\sigma}^{(\mu)}_{pq} \hat{\Lambda} (\alpha(t), \alpha^\dagger(t)) \nonumber\\
		+ \frac{i f}{\sqrt{N}} \sum_{\mu = 1}^{N} \left( \prod_{\nu \neq \mu} \hat{\rho}_\nu(t)  \right)
			\Biggl\{ \left( \rho^{(\mu)}_{ee}(t) \frac{\partial}{\partial \alpha^\dagger} \right) \hat{\sigma}^{(\mu)}_{eg} -
			\left( \rho^{(\mu)}_{ee}(t) \frac{\partial}{\partial \alpha} \right) \hat{\sigma}^{(\mu)}_{ge} 
		+
			\left[ \rho^{(\mu)}_{eg}(t) \frac{\partial}{\partial \alpha} - \rho^{(\mu)}_{ge}(t) \frac{\partial}{\partial \alpha^\dagger} \right] \hat{\sigma}^{(\mu)}_{gg} \Biggr\} \hat{\Lambda} (\alpha(t), \alpha^\dagger(t)) \nonumber\\
		\equiv \frac{i f}{\sqrt{N}} \sum_{\mu = 1}^{N} \left( \prod_{\nu \neq \mu} \hat{\rho}_\nu(t)  \right) \sum_{pq}
			\left[ \chi^{(\mu)}_{pq} (t) \frac{\partial}{\partial \alpha} + \chi^{\dagger(\mu)}_{pq} (t) \frac{\partial}{\partial \alpha^\dagger} \right]
			\hat{\sigma}^{(\mu)}_{pq} \hat{\Lambda} (\alpha(t), \alpha^\dagger(t)). \quad\quad\quad
\label{eqnA1:difference}
\end{eqnarray}

Although (\ref{eqnA1:difference}) looks quite bulky, it contains only the entangled pairs of single atom - field variables and do not involve correlators of higher orders.
As a result, one can attempt to reproduce them by switching from the deterministic to stochastic variables $\alpha(t)$, $\alpha^\dagger(t)$, $\rho_{pq}^{(\mu)} (t)$ and artificially introducing the appropriate noise terms (in the Ito's sense) into the equations (\ref{eqnA1:classic}):
\begin{gather}
	\left( \frac{d \alpha}{dt} \right)_\text{stoch} = \zeta (t), \quad \left( \frac{d \alpha^\dagger}{dt} \right)_\text{stoch} = \zeta^\dagger (t);
		\quad 	\left( \frac{d \rho_{pq}^{(\mu)}}{dt} \right)_\text{stoch} = \xi_{pq}^{(\mu)} (t),
\label{eqnA1:new-noise}
\end{gather}
\noindent with the following correlation properties:
\begin{gather}
	\langle \zeta(t) \xi^{(\mu)}_{pq} (t') \rangle = \kappa^{ (\mu)}_{pq} (t) \delta (t - t'), \quad
		\langle \zeta^\dagger (t) \xi^{ (\mu)}_{pq} (t') \rangle = \kappa^{\dagger (\mu)}_{pq} (t) \delta (t - t'),
\label{eqnA1:correlators}\\
	\text{all other correlators equal 0}. \nonumber
\end{gather}
\noindent Here $\langle ... \rangle$ denotes averaging over the stochastic sampling of the system's evolution.

As a result of (\ref{eqnA1:new-noise}), the total density matrix also becomes stochastic, and according to the Ito's lemma \cite{ito_stochastic_1944,protter_stochastic_2005}, its full time derivative acquires additional contribution
\begin{gather}
	\left\langle \sum_{\mu = 1}^{N}  \sum_{pq}
		\left[ \frac{\partial^2 \hat{\rho}_{\text{ans}}(t)}{\partial \rho^{(\mu)}_{pq} \partial \alpha}  \kappa^{ (\mu)}_{pq} (t)
		+ \frac{\partial^2 \hat{\rho}_{\text{ans}}(t)}{\partial \rho^{(\mu)}_{pq} \partial \alpha^\dagger}  \kappa^{\dagger (\mu)}_{pq} (t) \right] \right\rangle = \nonumber\\
	\left\langle \sum_{\mu = 1}^{N} \left( \prod_{\nu \neq \mu} \hat{\rho}_\nu(t)  \right) \sum_{pq}
	\left[ \kappa^{(\mu)}_{pq} (t) \frac{\partial}{\partial \alpha} + \kappa^{\dagger(\mu)}_{pq} (t) \frac{\partial}{\partial \alpha^\dagger} \right]
	\hat{\sigma}^{(\mu)}_{pq} \hat{\Lambda} (\alpha(t), \alpha^\dagger(t)) \right\rangle,
\label{eqnA1:rho-tot-noise}
\end{gather}
\noindent which structure coincides with that of expression (\ref{eqnA1:difference}). In order to obtain the full coincidence, we conclude that correlators (\ref{eqnA1:correlators}) should have the form
\begin{equation}
	\kappa^{ (\mu)}_{pq} (t) = \frac{i f}{\sqrt{N}} \chi^{(\mu)}_{pq} (t), \quad  \kappa^{\dagger(\mu)}_{pq} (t) = \frac{i f}{\sqrt{N}} \chi^{\dagger (\mu)}_{pq} (t),
\label{eqnA1:corr2}
\end{equation}
\noindent and can be easily sampled by means of the two independent complex valued gaussian white noise terms $F(t)$ and $F^\dagger (t)$  with mean 0 and variance 1 in the way as follows:
\begin{gather}
	\left( \frac{d \alpha}{dt} \right)_\text{stoch} = - i F (t), \quad \left( \frac{d \alpha^\dagger}{dt} \right)_\text{stoch} = i F^\dagger (t); \nonumber\\
	\left( \frac{d \rho_{pq}^{(\mu)}}{dt} \right)_\text{stoch} = \frac{f}{\sqrt{N}} \left[ - \chi^{(\mu)}_{pq}(t) F^* (t) 
		+ \chi^{\dagger (\mu)}_{pq}(t) F^{\dagger* } (t) \right].
\end{gather}
\noindent with the correlation properties:
\begin{gather}
	\langle F (t) F^* (t') \rangle = \delta(t- t'), \quad \langle F^\dagger (t) F^{\dagger *} (t') \rangle = \delta(t-t'), \nonumber\\
	\text{all other correlators equal 0}.
	\label{eqnA1:gammas}
\end{gather} 

Let us stress that numerical coefficients in front of the $\delta$-functions in (\ref{eqnA1:correlators}) can be splitted between the participating noise terms quite arbitrarily (this is also known as diffusion gauge \cite{deuar_gauge_2002,plimak_optimization_2001}).

Calculating the quantities $\chi^{(\mu)}_{pq}(t)$ and $\chi^{\dagger (\mu)}_{pq}(t)$ from (\ref{eqnA1:difference}) explicitly, we derive the system of SDEs, governing the quantum evolution of the TCM:

\begin{gather}
	\frac{d \alpha(t)}{dt}  = - i \omega_c \alpha(t) - i \frac{f}{\sqrt{N}} \sum_\nu \rho_{eg}^{(\nu)}(t)
		- i F(t), \nonumber \\
	\frac{d \alpha^\dagger(t)}{dt}  = ~~i \omega_c \alpha^\dagger(t) + i \frac{f}{\sqrt{N}} \sum_\nu \rho_{ge}^{(\nu)}(t)
		+ i F^\dagger(t), \nonumber \\
	\frac{d \rho_{ee}^{(\mu)}(t)}{dt}  = ~~i \frac{f}{\sqrt{N}} \left( \rho_{eg}^{(\mu)}(t) \alpha^\dagger(t) - \rho_{ge}^{(\mu)}(t) \alpha(t) \right)
	- \gamma \rho_{ee}^{(\mu)}(t)
		- \frac{f}{\sqrt{N}} \rho_{ee}^{(\mu)}(t)  \left[ \rho_{eg}^{(\mu)}(t) F^*(t) + \rho_{ge}^{(\mu)}(t) F^{\dagger *} (t)  \right],
\label{eqnA1:full}\\
	\frac{d \rho_{gg}^{(\mu)}(t)}{dt}  = - i \frac{f}{\sqrt{N}} \left( \rho_{eg}^{(\mu)}(t) \alpha^\dagger(t) - \rho_{ge}^{(\mu)}(t) \alpha(t) \right)
	+ \gamma \rho_{ee}^{(\mu)}(t)
		+ \frac{f}{\sqrt{N}} (1 - \rho_{gg}^{(\mu)}(t))  \left[ \rho_{eg}^{(\mu)}(t) F^*(t) + \rho_{ge}^{(\mu)}(t) F^{\dagger *} (t) \right], \nonumber\\
	\frac{d \rho_{eg}^{(\mu)}(t)}{dt}  = - i \omega_0 \rho_{eg}^{(\mu)}(t) + i \frac{f}{\sqrt{N}} \left( \rho_{ee}^{(\mu)}(t) - \rho_{gg}^{(\mu)}(t)\right) \alpha(t) - \frac{\gamma}{2} \rho_{eg}^{(\mu)}(t) \nonumber\\
		\quad\quad\quad\quad\quad\quad\quad\quad\quad
		 + \frac{f}{\sqrt{N}} \rho_{ee}^{(\mu)}(t) F^{\dagger *}(t)
		- \frac{f}{\sqrt{N}} \rho_{eg}^{(\mu)}(t)  \left[ \rho_{eg}^{(\mu)}(t) F^*(t) + \rho_{ge}^{(\mu)}(t) F^{\dagger *}(t)  \right] , \nonumber\\
	\frac{d \rho_{ge}^{(\mu)}(t)}{dt} = ~~i \omega_0 \rho_{ge}^{(\mu)}(t) - i \frac{f}{\sqrt{N}} \left( \rho_{ee}^{(\mu)}(t) - \rho_{gg}^{(\mu)}(t)\right) \alpha^\dagger(t) - \frac{\gamma}{2} \rho_{ge}^{(\mu)}(t) \nonumber\\
	\quad\quad\quad\quad\quad\quad\quad\quad\quad
		+ \frac{f}{\sqrt{N}} \rho_{ee}^{(\mu)}(t) F^{*}(t)
		- \frac{f}{\sqrt{N}} \rho_{ge}^{(\mu)}(t)  \left[ \rho_{eg}^{(\mu)}(t) F^*(t) + \rho_{ge}^{(\mu)}(t) F^{\dagger *} (t) \right] . \nonumber
\end{gather}

\subsection{Method II: Positive P-representation}
\label{sec:appA-II}

Alternative  way of deriving stochastic equations of motion consists in using the positive P-representation. Its main idea is based on deriving the Fokker-Planck equation for the specific phase-space distribution function, with the subsequent mapping of this equation to the coupled SDEs in Ito's form \cite{carmichael_statistical_1999,carmichael_statistical_2008}.

P-representation requires all quantum operators in the Hamiltonian to be normal ordered. Arranging this for the bosonic field operators is obvious, whereas its implementation with respect to the atomic $\hat{\sigma}$-projectors requires extra care.

It was shown in \cite{biedenharn1965quantum,sakurai_modern_1994,chuchurka_quantum_2023} that action of the $\hat{\sigma}$-projectors can be mimicked by a pair of  bosonic creation and annihliation operators defined as $\hat{\sigma}_{ij} = \hat{c}^\dagger_i \hat{c}_j$. The operator $\hat{c}_j$ destroys an atom in the state $\ket{j}$ and puts it into the  intermediate vacuum state $\ket{\o}$. Then operator $\hat{c}^\dagger_i$ acts onto the vacuum state and creates the atom in the state $\ket{i}$ (here for the matter of simplicity we omit the $\mu$-index, assuming that both operators and state vectors are referred to the same Hilbert space). Hence, each individual $\hat{\sigma}$-projector is already represented by normally ordered pair of  bosonic operators.

Following this idea, we fulfill the requirement of normal ordering in (\ref{H_TCM_proj}) by taking into account that particle and field operators commute with each other
and that $\hat{\sigma}$-projectors enter the Hamiltonian of TCM only linearly.

The next step consists in substituting the field operators and $\hat{\sigma}$-projectors by  analogous stochastic variables \cite{drummond_generalised_1980,chuchurka_quantum_2023}. These variables are complex valued functions of time representing the system's extended phase space (here the term ''extended'' means that we mathematically double the number of independent variables because each pair of initially conjugated bosonic operators is represented by  two independent complex numbers):
\begin{gather}
	\hat{a} \rightarrow \alpha (t), \quad \hat{a}^\dagger \rightarrow \beta (t) \equiv \alpha^\dagger (t); \quad \hat{\sigma}^{(\mu)}_{ij} \rightarrow \rho^{(\mu)}_{ji} (t),
\label{stoch_trans}
\end{gather}
\noindent where we substituted the $\hat{\sigma}$-projectors by the corresponding stochastic density matrix elements as described in \cite{chuchurka_quantum_2023}, thus skipping the intermediate substitution for the bosonic operators  $\hat{c}^{(\mu)}_i, \hat{c}^{\dagger (\nu)}_j \rightarrow C^{(\mu)}_i (t), C^{\dagger (\nu)}_j (t)$.  Here the dagger symbol in $\alpha^\dagger$ does not assume any mathematical operation and is used as a convenient notation. Let us stress that the indices of the stochastic density matrix elements $\rho$ should be taken in the inverse order with respect to the indices of the associated $\hat{\sigma}$-projector.

Applying the transformation rules (\ref{stoch_trans}) to (\ref{H_TCM_proj}), we derive the stochastic counterpart of the Hamiltonian of TCM:
\begin{equation}
	\hat{H} \rightarrow \mathcal{H} (t) = \frac{\hbar \omega_0}{2} \sum_{\mu = 1}^{N} \Bigl( \rho_{ee}^{(\mu)} (t) - \rho_{gg}^{(\mu)} (t) \Bigr) + \hbar \omega_c \alpha^\dagger (t) \alpha (t) + \frac{\hbar f}{\sqrt{N}}
		\sum_{\mu = 1}^{N} \Bigl( \rho_{ge}^{(\mu)}(t) \alpha(t) + \rho_{eg}^{(\mu)}(t) \alpha^\dagger(t) \Bigr),
\label{H_stoch}
\end{equation}
\noindent where all quantities are complex valued functions of time, and their ordering is inconsequential.

The general form of stochastic equations of motion \cite{chuchurka_quantum_2023} (incoherent processes will be included separately) for the field variables are
\begin{gather}
	\frac{d \alpha (t)}{dt} = - \frac{i}{\hbar} \frac{\partial \mathcal{H}}{\partial \alpha^\dagger} + \zeta(t), \quad\quad
		\frac{d \alpha^\dagger (t)}{dt} =  \frac{i}{\hbar} \frac{\partial \mathcal{H}}{\partial \alpha} + \zeta^\dagger (t),
\label{gen:a}
\end{gather}
\noindent and
\begin{equation}
	\frac{d \rho^{(\mu)}_{ij}   (t)}{dt} = \frac{i}{\hbar} \sum_{r = e,g} \Bigl[ \rho^{(\mu)}_{ir} (t) \frac{\partial \mathcal{H}}{\partial \rho^{(\mu)}_{jr} }-
		\frac{\partial \mathcal{H}}{\partial \rho^{(\mu)}_{ri}} \rho^{(\mu)}_{rj}(t) \Bigr] +
		\sum_{r = e,g} \Bigl[ \rho^{(\mu)}_{ir}(t) \xi^{\dagger(\mu)}_{rj} (t) + \rho^{(\mu)}_{rj}(t) \xi^{(\mu)}_{ir} (t)  \Bigr].
\label{gen:rho}
\end{equation}
The quantities $\zeta(t)$, $\zeta^{\dagger}(t)$ in (\ref{gen:a}) and $\xi_{ij}^{(\mu)}(t)$, $\xi_{ij}^{\dagger (\mu)}(t)$ in (\ref{gen:rho}) represent the stochastic terms, which by definition incorporate all quantum effects in the evolution of the considered system. These terms are not independent, but satisfy the following correlation properties \cite{chuchurka_quantum_2023}:
\begin{gather}
	\langle \zeta(t) \zeta (t') \rangle = -\frac{i}{\hbar} \frac{\partial^2 \mathcal{H}}{\partial \alpha^{\dagger 2}} \delta(t-t'), \quad
		\langle \zeta^\dagger(t) \zeta^\dagger (t') \rangle = \frac{i}{\hbar} \frac{\partial^2 \mathcal{H}}{\partial \alpha^2} \delta(t-t'), \nonumber\\
		\langle \zeta^\dagger(t) \zeta (t') \rangle = 0,
\label{corr:a-a}
\end{gather}
\begin{gather}
	\langle \xi^{\dagger (\mu)}_{ij}(t) \xi^{\dagger (\nu)}_{pq} (t') \rangle = -\langle \xi^{(\mu)}_{ij}(t) \xi^{(\nu)}_{pq} (t') =
		\frac{i}{\hbar} \frac{\partial^2 \mathcal{H}}{\partial \rho^{(\mu)}_{ji} \partial \rho^{(\nu)}_{qp}} \delta(t-t'), \nonumber \\
		\langle \xi^{\dagger (\mu)}_{ij}(t) \xi^{(\nu)}_{pq} (t') \rangle = 0,
\label{corr:ro-ro}
\end{gather}
\begin{gather}
	\langle \zeta(t) \xi^{(\mu)}_{ij} (t') \rangle = - \frac{i}{\hbar} \frac{\partial^2 \mathcal{H}}{\partial \alpha^\dagger \partial \rho^{(\mu)}_{ji}} \delta(t-t'), \quad
		\langle \zeta^\dagger(t) \xi^{\dagger (\mu)}_{ij} (t') \rangle = \frac{i}{\hbar} \frac{\partial^2 \mathcal{H}}{\partial \alpha \partial \rho^{(\mu)}_{ji}} \delta(t-t'), \nonumber \\
	\langle \zeta(t) \xi^{\dagger (\mu)}_{ij} (t') \rangle = \langle \zeta^\dagger(t) \xi^{(\mu)}_{ij} (t') \rangle = 0.
\label{corr:a-ro}
\end{gather}
\noindent where $\langle ... \rangle$ denotes averaging over the stochastic sampling of SDEs (\ref{gen:a})--(\ref{gen:rho}). 

We start with deterministic  part of the equations. Following the prescription of (\ref{gen:a})--(\ref{gen:rho}) and calculating the first-order derivatives from the stochastic Hamiltonian (\ref{H_stoch}), we obtain:
\begin{align}
	\left( \frac{d \alpha}{dt} \right)_\text{det} &= - i \omega_c \alpha (t) - i \frac{f}{\sqrt{N}} \sum_\nu \rho_{eg}^{(\nu)}(t), \nonumber \\
	\left( \frac{d \alpha^\dagger}{dt} \right)_\text{det} &= ~~i \omega_c \alpha^\dagger(t) + i \frac{f}{\sqrt{N}} \sum_\nu \rho_{ge}^{(\nu)}(t), \nonumber \\
	\left( \frac{d \rho_{ee}^{(\mu)}}{dt} \right)_\text{det} &= ~~i \frac{f}{\sqrt{N}} \left( \rho_{eg}^{(\mu)}(t) \alpha^\dagger(t) - \rho_{ge}^{(\mu)}(t) \alpha(t) \right), \nonumber\\
	\left( \frac{d \rho_{gg}^{(\mu)}}{dt} \right)_\text{det} &= - i \frac{f}{\sqrt{N}} \left( \rho_{eg}^{(\mu)}(t) \alpha^\dagger(t) - \rho_{ge}^{(\mu)}(t) \alpha \right),
\label{eqnA2:detpart}\\
	\left( \frac{d \rho_{eg}^{(\mu)}}{dt} \right)_\text{det} &= - i \omega_0 \rho_{eg}^{(\mu)}(t) + i \frac{f}{\sqrt{N}} \left( \rho_{ee}^{(\mu)}(t) - \rho_{gg}^{(\mu)}(t)\right) \alpha(t), \nonumber\\
	\left( \frac{d \rho_{ge}^{(\mu)}}{dt} \right)_\text{det} &= ~~i \omega_0 \rho_{ge}^{(\mu)}(t) - i \frac{f}{\sqrt{N}} \left( \rho_{ee}^{(\mu)}(t) - \rho_{gg}^{(\mu)}(t)\right) \alpha^\dagger(t).
\end{align}

In order to define the stochastic part of the equations, one should calculate the second-order derivatives from the stochastic Hamiltonian (\ref{H_stoch}) and substitute them into the correlators (\ref{corr:a-a})--(\ref{corr:a-ro}). In case of TCM, only a few of them have non-zero values:
\begin{align}
	\langle \zeta(t) \xi^{(\mu)}_{ge} (t') \rangle = - i \frac{f}{\sqrt{N}} \delta(t-t'), \quad
	\langle \zeta^\dagger(t) \xi^{\dagger (\mu)}_{eg} (t') \rangle = i \frac{f}{\sqrt{N}} \delta(t-t').
	\label{corr:nonzero}
\end{align}
\noindent Hence, the uncorrelated noise terms $\xi^{\dagger (\mu)}_{ee}(t)$, $\xi^{\dagger (\mu)}_{gg}(t)$, $\xi^{(\mu)}_{ee}(t)$, and $\xi^{(\mu)}_{gg}(t)$ do not contribute into the averages of the stochastic variables and can be omitted.

In order to mathematically reproduce the non-vanishing correlators (\ref{corr:nonzero}) in SDEs, one can model them by means of the two independent complex valued gaussian white noises $F(t)$ and $F^\dagger(t)$ with mean 0 and variance 1, which satisfy the following correlation properties:
\begin{gather}
	\langle F (t) F^* (t') \rangle = \delta(t-t'), \quad \langle F (t) F (t') \rangle = 0 = \langle F^* (t) F^* (t') \rangle, \nonumber \\
	\langle F^\dagger (t) F^{\dagger*} (t') \rangle = \delta(t-t'), \quad \langle F^\dagger (t) F^\dagger (t') \rangle = 0 = \langle F^{\dagger*} (t) F^{\dagger*} (t') \rangle, \\
	\text{any correlator between $F$ and $F^\dagger$ is zero}. \nonumber
\end{gather}

Splitting the numerical coefficient in front of the $\delta$-function in (\ref{corr:nonzero}) between the participating noise terms can be done quite arbitrarily (this is also known as diffusion gauge \cite{deuar_gauge_2002,plimak_optimization_2001}). Let us arrange them in the way as follows:
\begin{align}
	\zeta (t) = F (t), \quad \xi^{(\mu)}_{ge} (t) = - i \frac{f}{\sqrt{N}} F^* (t), \nonumber \\
	\zeta^\dagger(t) = F^\dagger (t), \quad \xi^{\dagger (\mu)}_{eg} (t)  = i \frac{f}{\sqrt{N}} F^{\dagger *} (t).
\label{eqnA2:noises}
\end{align}

The next step is to include the terms responsible for the incoherent processes. It was shown in \cite{chuchurka_quantum_2023} that the presence of  incoherent terms in the quantum master equation does not affect the part related to the closed system's evolution and leads to the inclusion of extra additive deterministic terms only:
\begin{equation}
	\Biggl( \frac{d \rho^{(\mu)}_{pq}}{dt} \Biggr)_\text{open} = \Biggl( \frac{d \rho^{(\mu)}_{pq}}{dt} \Biggr)_\text{closed} + \Biggl( \frac{d \rho^{(\mu)}_{pq}}{dt} \Biggr)_\text{bath},
\label{eq:eqwithbath}
\end{equation}
\noindent where in case of (\ref{eq:masterdiss})
\begin{align}
	\Biggl( \frac{d \rho^{(\mu)}_{ee}}{dt} \Biggr)_\text{bath} = - \gamma \rho^{(\mu)}_{ee}(t), \quad 
		\Biggl( \frac{d \rho^{(\mu)}_{eg}}{dt} \Biggr)_\text{bath} = - \frac{\gamma}{2} \rho^{(\mu)}_{eg}(t), \nonumber\\
	\Biggl( \frac{d \rho^{(\mu)}_{gg}}{dt} \Biggr)_\text{bath} = ~~\gamma \rho^{(\mu)}_{ee}(t), \quad 
		\Biggl( \frac{d \rho^{(\mu)}_{ge}}{dt} \Biggr)_\text{bath} = - \frac{\gamma}{2} \rho^{(\mu)}_{ge}(t).
\label{eq:addbath}
\end{align}

Bringing (\ref{eqnA2:detpart}), (\ref{eqnA2:noises}), and (\ref{eq:eqwithbath}) together, we derive the resulting set of SDEs:
\begin{align}
	\frac{d \alpha(t)}{dt}  &= - i \omega_c \alpha(t) - i \frac{f}{\sqrt{N}} \sum_\nu \rho_{eg}^{(\nu)}(t)
	- i F(t), \nonumber \\
	\frac{d \alpha^\dagger(t)}{dt}  &= ~~i \omega_c \alpha^\dagger(t) + i \frac{f}{\sqrt{N}} \sum_\nu \rho_{ge}^{(\nu)}(t)
	+ i F^\dagger(t), \nonumber \\
	\frac{d \rho_{ee}^{(\mu)}(t)}{dt}  &= ~~i \frac{f}{\sqrt{N}} \left( \rho_{eg}^{(\mu)}(t) \alpha^\dagger(t) - \rho_{ge}^{(\mu)}(t) \alpha(t) \right)
	- \gamma \rho_{ee}^{(\mu)}(t),
\label{eqnA2:full}\\
	\frac{d \rho_{gg}^{(\mu)}(t)}{dt}  &= - i \frac{f}{\sqrt{N}} \left( \rho_{eg}^{(\mu)}(t) \alpha^\dagger(t) - \rho_{ge}^{(\mu)}(t) \alpha(t) \right)
	+ \gamma \rho_{ee}^{(\mu)}(t)
	+ \frac{f}{\sqrt{N}}  \left[ \rho_{eg}^{(\mu)}(t) F^*(t) + \rho_{ge}^{(\mu)}(t) F^{\dagger *} (t) \right], \nonumber\\
	\frac{d \rho_{eg}^{(\mu)}(t)}{dt}  &= - i \omega_0 \rho_{eg}^{(\mu)}(t) + i \frac{f}{\sqrt{N}} \left( \rho_{ee}^{(\mu)}(t) - \rho_{gg}^{(\mu)}(t)\right) \alpha(t) - \frac{\gamma}{2} \rho_{eg}^{(\mu)}(t)
	+ \frac{f}{\sqrt{N}} \rho_{ee}^{(\mu)}(t) F^{\dagger *}(t), \nonumber\\
	\frac{d \rho_{ge}^{(\mu)}(t)}{dt} &= ~~i \omega_0 \rho_{ge}^{(\mu)}(t) - i \frac{f}{\sqrt{N}} \left( \rho_{ee}^{(\mu)}(t) - \rho_{gg}^{(\mu)}(t)\right) \alpha^\dagger(t) - \frac{\gamma}{2} \rho_{ge}^{(\mu)}(t)
	+ \frac{f}{\sqrt{N}} \rho_{ee}^{(\mu)}(t) F^{*}(t). \nonumber
\end{align}

At first glance, (\ref{eqnA2:full}) and (\ref{eqnA1:full}) look very different. However, such a conclusion is misleading because of the specific initial conditions incorporated into the equations (\ref{eqnA2:full}), which should be defined when carrying out the numerical simulation. For the field variables, this is trivial provided that the field starts either from vacuum or  coherent state because both are  eigenstates of the field annihilation operator. Choosing appropriate initial conditions for the atomic degrees of freedom is more delicate: the $\hat{c}$ and $\hat{c}^\dagger$ operators mentioned above act in a larger Hilbert space than the subspace of the atomic subsystem \cite{chuchurka_quantum_2023}. Furthermore, they appear always in pairs of $\hat{c}$ and $\hat{c}^\dagger$, since the atoms can not be annihilated but may only undergo transitions between atomic eigenstates (in such a representation they are the Fock-states for $\hat{c}^\dagger_i \hat{c}_i$). Acting with a single operator $\hat{c}$ would annihilate the atom creating non-physical vacuum state. This issue can be overcome by putting additional constraints on the atomic stochastic variables. Mathematically it can be introduced in the form of specific quasi-statistical properties incorporated into the initial conditions \cite{chuchurka_quantum_2023}:
\begin{align}
	\rho^{(\mu)}_{ij} (0) &= \eta^{(\mu)} \cdot \rho'^{(\mu)}_{ij} (0),  
	\label{eqn:eta_def}
\end{align} 
\noindent where $\rho'^{(\mu)}_{ij} (0)$ denote the usual initial occupancies and coherences of the density matrix; $\eta^{(\mu)}$  are time-independent special numbers mimicking the required constraints by the following correlators (here the notation $\langle ... \rangle_\eta$ denotes additional averaging over the $\eta$-numbers sampling, which is complementary to the stochastic sampling):
\begin{align}
	\biggl\langle \left(\eta^{(\mu)} \right)^m \biggr\rangle_\eta = 
	\begin{cases}
		1, \quad \text{ if } m = 0 \text{ or } 1 \\
		0, \quad \text{ if } m > 1
	\end{cases}.
	\label{eqn:eta_properties}
\end{align}

Although the system of equations (\ref{eqnA2:full})--(\ref{eqn:eta_properties}) is now full and self-consistent, sampling the initial conditions by means of $\eta$-numbers and averaging over them is very inconvenient. Instead, the effect of the $\eta$-numbers can be treated by extra noise terms introduced directly into the equations (\ref{eqnA2:full}) (see Appendix \ref{sec:appB} for details): 
\begin{equation}
	\frac{d \rho^{(\mu)}_{pq}(t)}{dt} = \dotso -  \rho^{(\mu)}_{pq}(t) \frac{f}{\sqrt{N}} \Bigl[  \rho^{(\mu)}_{eg}(t) F^* (t) + \rho^{(\mu)}_{ge}(t) F^{\dagger *} (t) \Bigr].
	\label{eqn:eta_update}
\end{equation}

Let us stress that the field equations are not affected by the transformation (\ref{eqn:eta_def}). Including the terms (\ref{eqn:eta_update}) in (\ref{eqnA2:full}) results in the total coincidence of (\ref{eqnA2:full}) with (\ref{eqnA1:full}). 

\subsection{Additional rearrangements in stochastic equations}
\label{sec:appA-III}

First of all, one can benefit from the full permutational symmetry of the atomic subsystem within the TCM, which results in all $\hat{\rho}_\mu$ having the same matrix elements $\rho^{(\mu)}_{pq} (t) \, \rightarrow \, \rho_{pq} (t)$ and removal of the $\mu$-index from the equations. It is also easy to show that the sum of stochastic variables $\rho_{ee} (t)$ and $\rho_{gg} (t)$  is an integral of motion for each individual trajectory in the considered case (see Appendix \ref{sec:appD}):
\begin{equation}
	\rho_{ee} (t) + \rho_{gg} (t) = 1, 
\end{equation}
\noindent so that the system can be reduced to five coupled equations of motion:
\begin{gather}
	\frac{d \alpha(t)}{dt}  = - i \omega_c \alpha(t) - i f\sqrt{N} \rho_{eg}(t)- i F(t), \nonumber \\
	\frac{d \alpha^\dagger(t)}{dt}  = ~~i \omega_c \alpha^\dagger(t) + i f\sqrt{N} \rho_{ge}(t)	+ i F^\dagger(t), \nonumber \\
	\frac{d \rho_{ee}(t)}{dt}  = ~~i \frac{f}{\sqrt{N}} \left( \rho_{eg}(t) \alpha^\dagger(t) - \rho_{ge}(t) \alpha(t) \right)
	- \gamma \rho_{ee}(t)
	- \frac{f}{\sqrt{N}} \rho_{ee}(t)  \left[ \rho_{eg}(t) F^*(t) + \rho_{ge}(t) F^{\dagger *} (t)  \right],
\label{eqnA3:simpl01}\\
	\frac{d \rho_{eg}(t)}{dt}  = - i \omega_0 \rho_{eg}(t) + i \frac{f}{\sqrt{N}} \left( 2 \rho_{ee}^{(\mu)}(t) - 1\right) \alpha(t) - \frac{\gamma}{2} \rho_{eg}(t) \nonumber\\
	\quad\quad\quad\quad\quad\quad\quad\quad\quad\quad
	+ \frac{f}{\sqrt{N}} \rho_{ee}(t) F^{\dagger *}(t)
	- \frac{f}{\sqrt{N}} \rho_{eg}(t)  \left[ \rho_{eg}(t) F^*(t) + \rho_{ge}(t) F^{\dagger *}(t)  \right] , \nonumber\\
	\frac{d \rho_{ge}(t)}{dt} = ~~i \omega_0 \rho_{ge}(t) - i \frac{f}{\sqrt{N}} \left( 2 \rho_{ee}^{(\mu)}(t) - 1\right) \alpha^\dagger(t) - \frac{\gamma}{2} \rho_{ge}(t) \nonumber\\
	\quad\quad\quad\quad\quad\quad\quad\quad\quad\quad
	+ \frac{f}{\sqrt{N}} \rho_{ee}(t) F^{*}(t)
	- \frac{f}{\sqrt{N}} \rho_{ge}(t)  \left[ \rho_{eg}(t) F^*(t) + \rho_{ge}(t) F^{\dagger *} (t) \right] . \nonumber
\end{gather}

We would like to point out two features of the SDEs (\ref{eqnA3:simpl01}), which might badly affect results of the numerical simulation: (i) all atomic equations contain multiplicative non-linear noise terms, eventually resulting in large and quick dispersion of the stochastic trajectories. This may lead to faster appearance of the "runaway" trajectories through the channels described in Section \ref{sec:3-2}. (ii) the field equations contain additive noise terms, which are not damped during the simulation (even when strong dissipation $\gamma$ is introduced). As a result, these terms also drive the increase of dispersion of the stochastic trajectories leading to the same issues as in the former case. Fortunately, these features can be removed by introducing the new independent complex valued white noise term $S(t)$ and taking into account that all quantum effects are embedded into the correlators between the stochastic variables. In order to show this explicitly, let us write down the non-zero correlators between the stochastic variables, coming from equations (\ref{eqnA3:simpl01}) and using the same notations for the stochastic terms as in (\ref{eqnA1:new-noise}) and (\ref{gen:a})--(\ref{gen:rho}):
\begin{align}
	&\langle \zeta(t ) \xi_{ee} (t') \rangle = \frac{i f}{\sqrt{N}} \rho_{ee}(t)\rho_{eg}(t) \delta(t - t'), \quad
	&&\langle \zeta^{\dagger}(t) \xi_{ee} (t') \rangle = - \frac{i f}{\sqrt{N}} \rho_{ee} (t) \rho_{ge}(t) \delta(t - t'), \nonumber\\
	&\langle \zeta(t ) \xi_{eg} (t') \rangle = \frac{i f}{\sqrt{N}} \rho_{eg}(t) \rho_{eg}(t) \delta(t - t'), \quad
	&&\langle \zeta^{\dagger}(t) \xi_{eg} (t') \rangle = \frac{i f}{\sqrt{N}} \bigl(\rho_{ee}(t) - \rho_{eg}(t) \rho_{ge}(t) \bigr) \delta(t - t'), 
	\label{eq:appC:correlators} \\
	&\langle \zeta(t ) \xi_{ge} (t') \rangle = - \frac{i f}{\sqrt{N}} \bigl(\rho_{ee}(t) - \rho_{ge} (t) \rho_{eg} (t) \bigr) \delta(t - t'), \quad
	&&\langle \zeta^{\dagger}(t) \xi_{ge} (t') \rangle = - \frac{i f}{\sqrt{N}} \rho_{ge}(t) \rho_{ge}(t) \delta(t - t'), \nonumber
\end{align}

It is easy to show that the same correlators can be reproduced by the following stochastic terms:
\begin{gather}
	\zeta(t) = - i \sqrt{\rho_{ee}(t)} F(t) -i \rho_{eg}(t) S(t), \quad\quad
		\zeta^\dagger(t) = i \sqrt{\rho_{ee}(t)} F^\dagger(t) +  i \rho_{ge} (t) S(t); \nonumber\\ 
	\xi_{ee}(t) = - \frac{f}{\sqrt{N}} \rho_{ee}(t) S^*(t), 
\label{eqnA3:newnoises}\\
	\xi_{eg} (t) = \frac{f}{\sqrt{N}} \sqrt{\rho_{ee}(t)} F^{\dagger *}(t) - \frac{f}{\sqrt{N}} \rho_{eg}(t) S^*(t), \quad\quad
		\xi_{ge} (t) = \frac{f}{\sqrt{N}} \sqrt{\rho_{ee} (t)} F^*(t) - \frac{f}{\sqrt{N}} \rho_{ge}(t) S^*(t). \nonumber 
\end{gather}
\noindent where we took into account
\begin{equation}
	\langle S(t)S^*(t') =\delta(t-t')\rangle.
\end{equation}

Substituting the stochastic terms in (\ref{eqnA3:simpl01}) by (\ref{eqnA3:newnoises}), we derive (\ref{eqn:final}).

\subsection{Equations for the numerical simulation}
\label{sec:appA-IV}

Within numerical simulations we consider the simplified case of the exact resonance between the field frequency and the atomic transition frequency $\omega_0 = \omega_c$.  This allows us to remove the quickly oscillating terms from the equations (\ref{eqn:final}) by means of the following transformation of the variables:
\begin{gather}
	\alpha(t) \, \rightarrow \, \alpha(t) e^{-i \omega_0 t},  \quad \alpha^\dagger (t) \, \rightarrow \, \alpha^\dagger(t) e^{i \omega_0 t}, \nonumber\\
	\rho_{eg} (t) \, \rightarrow \, \rho_{eg} (t) e^{-i \omega_0 t}, \quad \rho_{ge} (t) \, \rightarrow \, \rho_{ge} (t) e^{i \omega_0 t}.
\end{gather}

It is also convenient to introduce the normalized field variables
\begin{equation}
	A(t) = \frac{\alpha(t)}{\sqrt{N}}, \qquad A^\dagger(t) = \frac{\alpha^\dagger(t)}{\sqrt{N}},
\end{equation}
\noindent and use the dimensionless time $\tau = f t$, thus making the obtained results independent of $f$ (assuming renormalization of the decay constant $\gamma \, \rightarrow \, \frac{\gamma}{f}$). Doing this, one should take into account the following property of the delta function:
\begin{equation}
	\delta(t-t') = \delta\Bigl(\frac{\tau - \tau'}{f} \Bigr) = f \cdot \delta(\tau - \tau').
\end{equation}

As a result, all correlators in (\ref{eqn:gamfinal}) gain an extra $f$-factor, which should be included into the stochastic part of the equations. Applying diffusion gauge, we put this factor into the stochastic part of the field equations and do not change the stochastic part of the atomic equations. Finally, we derive:
\begin{gather}
	\frac{d A(\tau)}{d\tau}  = - i \rho_{eg}(\tau) - i \sqrt{\rho_{ee}(\tau)} \frac{F(\tau)}{\sqrt{N}} -i \rho_{eg}(\tau) \frac{S(\tau)}{\sqrt{N}}, \nonumber \\
	\frac{d A^\dagger(\tau)}{d\tau}  = ~~ i \rho_{ge}(t) + i \sqrt{\rho_{ee}(\tau)} \frac{F^\dagger(\tau)}{\sqrt{N}} +  i \rho_{ge} (\tau) \frac{S(\tau)}{\sqrt{N}}, \nonumber \\
	\frac{d \rho_{ee}(\tau)}{d\tau}  = - \frac{\gamma}{f} \rho_{ee}(\tau) + i \Bigl( \rho_{eg}(\tau) A^\dagger(\tau) - \rho_{ge}(\tau) A(\tau) \Bigr)
	- \rho_{ee}(\tau) \frac{S^*(\tau)}{\sqrt{N}},
	\label{eqn:numerics}\\
	\frac{d \rho_{eg}(\tau)}{d\tau}  = - \frac{\gamma}{2f} \rho_{eg}(\tau) + i \Bigl( 2 \rho_{ee}(\tau) - 1\Bigr) A(\tau) 
	+  \sqrt{\rho_{ee}(\tau)} \frac{F^{\dagger *}(\tau)}{\sqrt{N}} - \rho_{eg}(\tau) \frac{S^*(\tau)}{\sqrt{N}}, \nonumber\\
	\frac{d \rho_{ge}(\tau)}{d\tau} = - \frac{\gamma}{2f} \rho_{ge}(\tau)  - i \Bigl( 2 \rho_{ee}(\tau) - 1\Bigr) A^\dagger(\tau)  
	+ \sqrt{\rho_{ee} (\tau)} \frac{F^*(\tau)}{\sqrt{N}} - \rho_{ge}(\tau) \frac{S^*(\tau)}{\sqrt{N}}, \nonumber
\end{gather}
\noindent where $F(\tau)$, $F^\dagger(\tau)$, and $S(\tau)$ are the independent complex valued gaussian white noise terms, satisfying the correlation properties:
\begin{equation}
	\langle F(\tau) F^* (\tau') \rangle = \delta(\tau- \tau'), \quad \langle F^\dagger (\tau) F^{\dagger*} (\tau') \rangle = \delta(\tau- \tau'),
		\quad \langle S (\tau) S^* (\tau') \rangle = \delta(\tau- \tau').
\end{equation}

\section{Representing $\eta$-numbers as extra noise terms }
\label{sec:appB}

In order to take out the $\eta$-numbers from the atomic variables explicitly we start with the so-called characteristic function \cite{chuchurka_quantum_2023}, which stores all necessary information for calculating the expectation value of any observable:
\begin{equation}
	\chi (\vec{\lambda}, t) = \langle e^{\vec{\lambda} \cdot \vec{x} (t)} \rangle.
\label{eq:chidef}
\end{equation}
\noindent Here $\vec{x} (t)$ describes an arbitrary stochastic process and is governed by following set of SDEs:
\begin{equation}
	\frac{d \vec{x} (t)}{dt} = \vec{\Mu} (\vec{x},t) + \vec{\xi} (\vec{x},t),
\end{equation}
\noindent where $\vec{\Mu} (\vec{x},t)$ represents the deterministic (drift) terms and $\vec{\xi} (\vec{x},t)$ describes the stochastic (noise) terms.

Following definition (\ref{eq:chidef}), the average of any variable  $\mathcal{O} \left(\vec{x}(t)\right)$ can be calculated as:
\begin{equation}
	\langle \mathcal{O} \left(\vec{x}(t)\right) \rangle = \mathcal{O} \Bigl( \frac{\partial}{\partial \vec{\lambda}} \Bigr) \chi (\vec{\lambda}, t) \Bigr\rvert_{\vec{\lambda} = 0}.
\label{eq:chiaver}
\end{equation}
One of the striking features of the stochastic formalism consists in the possibility to construct another set of SDEs written in terms of the new stochastic variables $\vec{x}'(t)$, which describe the same physical system. Transition to the new variables is accompanied by the change of the charactertic function $\chi'(\vec{\lambda},t)$. However, as it was shown in \cite{chuchurka_quantum_2023},  certain relationships between the old and new characteristic functions should still hold:
\begin{gather}
	\chi (\vec{\lambda}, t) = \chi'(\vec{\lambda},t), \label{eq:genchicond1}\\
	\frac{\partial }{\partial t} \chi (\vec{\lambda}, t) = \frac{\partial }{\partial t} \chi'(\vec{\lambda},t).
	\label{eq:genchicond2}
\end{gather}

In particular, equations (\ref{eq:genchicond1})--(\ref{eq:genchicond2}) can be used directly to extract the $\eta$-numbers from the atomic variables and obtain $\eta$-free SDEs. In order to do that, let us introduce the characteristic function for the two cases of the considered model: (i) when the atomic variables implicitly contain the $\eta$-numbers through the initial conditions, and (ii) when the presence of the $\eta$-numbers is written explicitly as a factor in front of the $\eta$-free variables:
\begin{gather}
	\chi (\vec{\lambda}, t) = \langle e^{\sum_{\mu}\sum_{pq} \lambda^{(\mu)}_{pq} \rho^{(\mu)}_{pq} + \lambda \alpha + \lambda^\dagger \alpha^\dagger } \rangle \label{eq:chiold}, \\
	\chi' (\vec{\lambda}, t) = \langle e^{\sum_{\mu} \eta^{(\mu)}  \sum_{pq} \lambda^{(\mu)}_{pq} \rho'^{(\mu)}_{pq} + \lambda \alpha + \lambda^\dagger \alpha^\dagger } \rangle \label{eq:chinew}.
\end{gather}
\noindent Here $\lambda^{(\mu)}_{pq}$, $\lambda$ and $\lambda^\dagger$ are linearly independent quantities and notation $\lambda^\dagger$ is used for convenience purposes only (for any quantity here and below the dagger symbol does not presume any mathematical operation). We also assume that all quantities here and below except for $\lambda^{(\mu)}_{pq}$, $\lambda$, $\lambda^\dagger$, and $\eta^{(\mu)}$ are time-dependent.

Calculating the partial time derivative from (\ref{eq:chiold}) and using the Ito's lemma \cite{ito_stochastic_1944,protter_stochastic_2005}, one can deduce:
\begin{eqnarray}
	\frac{\partial }{\partial t} \chi (\vec{\lambda}, t) = \Biggl\langle \left\{ \lambda \Mu_\alpha + \lambda^\dagger \Mu_{\alpha^\dagger} + \sum_{\mu} \sum_{pq} \lambda^{(\mu)}_{pq}\Mu^{(\mu)}_{pq} + 
		+ \sum_{\mu}\sum_{pq} \lambda^{(\mu)}_{pq} \left( \lambda \text{C}^{(\mu)}_{pq,\alpha} + \lambda^\dagger \text{C}^{(\mu)}_{pq,\alpha^\dagger} \right)\right\} \times \nonumber\\
		\times e^{\sum_{\mu}\sum_{pq} \lambda^{(\mu)}_{pq} \rho^{(\mu)}_{pq} + \lambda \alpha + \lambda^\dagger \alpha^\dagger } \Biggr\rangle,
\label{eq:ddtchi1}	
\end{eqnarray}
\noindent where $\Mu_\alpha$, $\Mu_{\alpha^\dagger}$, and $\Mu^{(\mu)}_{pq}$ are the short notations for the deterministic part of the field and atomic SDEs; $\text{C}^{(\mu)}_{pq,\alpha}$ and $\text{C}^{(\mu)}_{pq,\alpha^\dagger}$ are the coefficients in the correlators between the stochastic parts of SDEs for atomic and field variables (this coefficients do not include the $\delta$-functions):
\begin{equation}
	\langle \xi^{(\mu)}_{pq} (t) \zeta (t') \rangle = \text{C}^{(\mu)}_{pq,\alpha} \delta(t - t'), \quad \langle \xi^{(\mu)}_{pq} (t) \zeta^\dagger(t') \rangle = \text{C}^{(\mu)}_{pq,\alpha^\dagger} \delta(t - t').
\end{equation}
\noindent  They can be written explicitly by addressing the equations (\ref{eqnA2:full}).

Repeating the same procedure for (\ref{eq:chinew}), we obtain:
\begin{eqnarray}
	\frac{\partial }{\partial t} \chi' (\vec{\lambda}, t) = \Biggl\langle \left\{ \lambda \Mu'_\alpha + \lambda^\dagger \Mu'_{\alpha^\dagger}
	+ \sum_{\mu} \eta^{(\mu)} \sum_{pq} \lambda^{(\mu)}_{pq}\Mu'^{(\mu)}_{pq} +  \sum_{\mu} \eta^{(\mu)} \sum_{pq} \lambda^{(\mu)}_{pq} \left( \lambda \text{C}'^{(\mu)}_{pq,\alpha} + \lambda^\dagger \text{C}'^{(\mu)}_{pq,\alpha^\dagger} \right)\right\} \times \nonumber\\
		\times e^{\sum_{\mu} \eta^{(\mu)}  \sum_{pq} \lambda^{(\mu)}_{pq} \rho'^{(\mu)}_{pq} + \lambda \alpha + \lambda^\dagger \alpha^\dagger } \Biggr\rangle.
\label{eq:ddtchi2}
\end{eqnarray}

Following the idea of equation (\ref{eq:chiaver}), one can extract the function written in braces of (\ref{eq:ddtchi1}) as:
\begin{equation}
	\Biggl\langle \mathcal{F} \left(\left\{\rho^{(\nu)}_{rs} \right\},\alpha,\alpha^\dagger\right) e^{\sum_{\mu}\sum_{pq} \lambda^{(\mu)}_{pq} \rho^{(\mu)}_{pq} + \lambda \alpha + \lambda^\dagger \alpha^\dagger } \Biggr\rangle =
		\mathcal{F} \left( \left\{\frac{\partial}{\partial \lambda^{(\nu)}_{rs}} \right\},\frac{\partial}{\partial \lambda},\frac{\partial}{\partial \lambda^\dagger} \right) \chi (\vec{\lambda}, t).
\label{eq:chiF1}
\end{equation}
Then taking into account condition (\ref{eq:genchicond1}) and substituting the function $\chi (\vec{\lambda},t)$ by the expression (\ref{eq:chinew}), we derive:
\begin{eqnarray}
	\mathcal{F} \left( \left\{\frac{\partial}{\partial \lambda^{(\nu)}_{rs}} \right\},\frac{\partial}{\partial \lambda},\frac{\partial}{\partial \lambda^\dagger} \right) \chi (\vec{\lambda}, t) =
		\mathcal{F} \left( \left\{\frac{\partial}{\partial \lambda^{(\nu)}_{rs}} \right\},\frac{\partial}{\partial \lambda},\frac{\partial}{\partial \lambda^\dagger} \right) \chi' (\vec{\lambda}, t) = \nonumber \\
	= \Biggl\langle \mathcal{F} \left(\left\{\eta^{(\nu)} \cdot \rho'^{(\nu)}_{pq} \right\},\alpha,\alpha^\dagger\right) e^{\sum_{\mu} \eta^{(\mu)}  \sum_{pq} \lambda^{(\mu)}_{pq} \rho'^{(\mu)}_{pq} + \lambda \alpha + \lambda^\dagger \alpha^\dagger } \Biggr\rangle.
\label{eq:chiF2}
\end{eqnarray}

Applying the rule (\ref{eq:chiF2}) to (\ref{eq:ddtchi1}), we derive:
\begin{eqnarray}
	\frac{\partial \chi}{\partial t}  = 
	\Biggl\langle \Biggl\{ \lambda \left[- i \omega_c \alpha - i \frac{f}{\sqrt{N}} \sum_{\mu} \eta^{(\mu)} \rho'^{(\mu)}_{eg} \right] +
		\lambda^\dagger \left[ i \omega_c \alpha^\dagger + i \frac{f}{\sqrt{N}} \sum_{\mu} \eta^{(\mu)} \rho'^{(\mu)}_{ge} \right] + \nonumber\\
	\sum_{\mu} \eta^{(\mu)} \sum_{pq} \lambda^{(\mu)}_{pq}\Mu^{(\mu)}_{pq} +
		\sum_{\mu} \eta^{(\mu)} \sum_{pq} \lambda^{(\mu)}_{pq} \bigl[ \lambda \text{C}^{(\mu)}_{pq,\alpha} + \lambda^\dagger \text{C}^{(\mu)}_{pq,\alpha^\dagger}  \bigr]\Biggr\} 
		e^{\sum_{\mu} \eta^{(\mu)}  \sum_{pq} \lambda^{(\mu)}_{pq} \rho'^{(\mu)}_{pq} + \lambda \alpha + \lambda^\dagger \alpha^\dagger } \Biggr\rangle,
	\label{eq:etadiff1expl}
\end{eqnarray}
\noindent where we substituted explicitly
\begin{gather}
	\Mu_\alpha \left(\left\{\eta^{(\nu)} \cdot \rho'^{(\nu)}_{pq} \right\},\alpha,\alpha^\dagger\right) = 
	- i \omega_c \alpha - i \frac{f}{\sqrt{N}} \sum_{\mu} \eta^{(\mu)} \rho^{(\mu)}_{eg}, \nonumber\\
	\Mu_{\alpha^\dagger} \left(\left\{\eta^{(\nu)} \cdot \rho'^{(\nu)}_{pq} \right\},\alpha,\alpha^\dagger\right) = 
	i \omega_c \alpha^\dagger + i \frac{f}{\sqrt{N}} \sum_{\mu} \eta^{(\mu)} \rho^{(\mu)}_{ge},
\end{gather}
and took into account that
\begin{equation}
	\Mu^{(\mu)}_{pq} \left(\eta^{(\mu)} \cdot \rho'^{(\mu)}_{pq},\alpha, \alpha^\dagger\right) = 
	\eta^{(\mu)} \Mu^{(\mu)}_{pq} \left(\rho'^{(\mu)}_{pq},\alpha, \alpha^\dagger\right).
	\label{eqnB:etaM}
\end{equation}
Identity (\ref{eqnB:etaM}) also holds for the quantities $\text{C}^{(\mu)}_{pq,\alpha}$ and $\text{C}^{(\mu)}_{pq,\alpha^\dagger}$.

Subtracting (\ref{eq:etadiff1expl}) from (\ref{eq:ddtchi2}), we obtain:
\begin{eqnarray}
	\frac{\partial \chi'}{\partial t}  - \frac{\partial \chi}{\partial t}  = 
		\Biggl\langle \Biggl\{ \sum_{\mu} \eta^{(\mu)} \sum_{pq} \lambda^{(\mu)}_{pq}\left(\Mu'^{(\mu)}_{pq} - \Mu^{(\mu)}_{pq}\right) + \nonumber\\
	+ \lambda \left[\Mu'_\alpha + i \omega_c \alpha + i \frac{f}{\sqrt{N}} \sum_{\mu} \eta^{(\mu)} \rho'^{(\mu)}_{eg} \right] +
		\lambda^\dagger \left[\Mu'_{\alpha^\dagger} - i \omega_c \alpha^\dagger - i \frac{f}{\sqrt{N}} \sum_{\mu} \eta^{(\mu)} \rho'^{(\mu)}_{ge} \right] + \nonumber \\
		+ \sum_{\mu} \eta^{(\mu)} \sum_{pq} \lambda^{(\mu)}_{pq} \bigl[ \lambda (\text{C}'^{(\mu)}_{pq,\alpha} -\text{C}^{(\mu)}_{pq,\alpha}) + \lambda^\dagger (\text{C}'^{(\mu)}_{pq,\alpha^\dagger} - \text{C}^{(\mu)}_{pq,\alpha^\dagger})  \bigr]\Biggr\} 
			e^{\sum_{\mu} \eta^{(\mu)}  \sum_{pq} \lambda^{(\mu)}_{pq} \rho'^{(\mu)}_{pq} + \lambda \alpha + \lambda^\dagger \alpha^\dagger } \Biggr\rangle.
\label{eq:etadiff1}
\end{eqnarray}

Now one can introduce concrete desired relationships between the old and new drift terms as well as between the old and new correlation coefficients. In the considered case it is convenient to assume that only the noise terms of the atomic equations are affected by the transformation, and all other terms keep the same functional dependencies, i.e.
\begin{gather}
	\Mu'^{(\mu)}_{pq} (\vec{x}',t) = \Mu^{(\mu)}_{pq} (\vec{x}',t), \quad \Mu'_\alpha (\vec{x}',t) = \Mu_\alpha (\vec{x}',t), \quad \Mu'_{\alpha^\dagger} (\vec{x}',t) = \Mu_{\alpha^\dagger} (\vec{x}',t), \nonumber \\
	B'_{\alpha} (\vec{x}',t)= B_{\alpha} (\vec{x}',t), \quad B'_{\alpha^\dagger} (\vec{x}',t) = B_{\alpha^\dagger} (\vec{x}',t), 
\label{eq:etafunccond}
\end{gather}
\noindent where $B_\alpha$ and $B_{\alpha^\dagger}$ are taken from the noise terms of the field equations and contribute to the correlation coefficients as 
\begin{equation}
	\text{C}^{(\mu)}_{pq,\alpha} = B^{(\mu)}_{pq,\alpha} B_\alpha, \quad \text{C}^{(\mu)}_{pq,\alpha^\dagger} = B^{(\mu)}_{pq,\alpha^\dagger} B_{\alpha^\dagger}.
\label{eq:etacorrcond}
\end{equation}
Applying (\ref{eq:etafunccond})--(\ref{eq:etacorrcond}), equation (\ref{eq:etadiff1}) is much simplified:
\begin{eqnarray}
	\frac{\partial \chi'}{\partial t}  - \frac{\partial \chi}{\partial t}  = 
	\Biggl\langle \Biggl\{ - \frac{ i f}{\sqrt{N}} \lambda \sum_{\mu} (1 - \eta^{(\mu)}) \rho'^{(\mu)}_{eg}  +
		\frac{ i f}{\sqrt{N}} \lambda^\dagger \sum_{\mu} (1- \eta^{(\mu)}) \rho'^{(\mu)}_{ge} + \nonumber \\
	+ \sum_{\mu} \eta^{(\mu)} \sum_{pq} \lambda^{(\mu)}_{pq} \bigl[ \lambda B_\alpha (\text{B}'^{(\mu)}_{pq,\alpha} -\text{B}^{(\mu)}_{pq,\alpha})
		+ \lambda^\dagger B_{\alpha^\dagger}(\text{B}'^{(\mu)}_{pq,\alpha^\dagger} - \text{B}^{(\mu)}_{pq,\alpha^\dagger})  \bigr]\Biggr\} 
	e^{\sum_{\mu} \eta^{(\mu)}  \sum_{pq} \lambda^{(\mu)}_{pq} \rho'^{(\mu)}_{pq} + \lambda \alpha + \lambda^\dagger \alpha^\dagger } \Biggr\rangle.
	\label{eqnB:chidiff-simpl}
\end{eqnarray}

Finally, following the identity derived in \cite{chuchurka_quantum_2023}
\begin{equation}
	\Bigl\langle  (1 - \eta^{(\nu)}) e^{y \eta^{(\nu)})} \Bigl\rangle_{\eta^{(\nu)}} = \Bigl\langle y \eta^{(\nu)} e^{y \eta^{(\nu)})} \Bigr\rangle_{\eta^{(\nu)}},
\end{equation}
\noindent we reduce (\ref{eq:etadiff1}) to:
\begin{eqnarray}
	\frac{\partial \chi'}{\partial t}  - \frac{\partial \chi}{\partial t}  = \nonumber\\
		\Biggl\langle \Biggl\{ \sum_{\mu} \eta^{(\mu)}\sum_{pq} \lambda^{(\mu)}_{pq} \biggl[ \lambda \left( B_\alpha ( B'^{(\mu)}_{pq,\alpha} - B^{(\mu)}_{pq,\alpha}) - \frac{i f}{\sqrt{N} }  \rho'^{(\mu)}_{pq} \rho'^{(\mu)}_{eg}\right) +
		\lambda^\dagger \left( B_{\alpha^\dagger} ( B'^{(\mu)}_{pq,\alpha^\dagger} - B^{(\mu)}_{pq,\alpha^\dagger}) + \frac{i f}{\sqrt{N} }  \rho'^{(\mu)}_{pq} \rho'^{(\mu)}_{ge}  \right) \biggr] \times \nonumber\\
	\times e^{\sum_{\mu} \eta^{(\mu)}  \sum_{pq} \lambda^{(\mu)}_{pq} \rho'^{(\mu)}_{pq} + \lambda \alpha + \lambda^\dagger \alpha^\dagger } \Biggr\rangle = 0. \nonumber\\
\label{eqnB:chidiff2}
\end{eqnarray}

Substituting the functions $B_\alpha$, and $B_{\alpha^\dagger}$ explicitly, one can obtain the following conditions:
\begin{gather}
	B'^{(\mu)}_{pq,\alpha} - B^{(\mu)}_{pq,\alpha} \equiv \Delta B^{(\mu)}_{pq,\alpha} = - \frac{f}{\sqrt{N}} \rho'^{(\mu)}_{pq} \rho'^{(\mu)}_{eg}, \nonumber\\
	B'^{(\mu)}_{pq,\alpha^\dagger} - B^{(\mu)}_{pq,\alpha^\dagger} \equiv \Delta B^{(\mu)}_{pq,\alpha^\dagger} = - \frac{f}{\sqrt{N}} \rho'^{(\mu)}_{pq} \rho'^{(\mu)}_{ge}.
\label{eq:finB1}
\end{gather}

Hence, in terms of the $\eta$-free variables all atomic equations should be updated in the way as follows:
\begin{equation}
	\frac{d \rho'^{(\mu)}_{pq} (t)}{dt} \, \rightarrow \, \frac{d \rho^{(\mu)}_{pq}(t)}{dt} \Bigr\rvert_{\rho^{(\mu)}_{pq} = \rho'^{(\mu)}_{pq}} 
		-  \rho'^{(\mu)}_{pq} (t) \frac{f}{\sqrt{N}} \Bigl[  \rho'^{(\mu)}_{eg}(t) F^* (t) + \rho'^{(\mu)}_{ge} (t) F^{\dagger *} (t) \Bigr].
\end{equation}

Without the loss of generality, we omit the prime symbol in equations (\ref{eqn:eta_update}).

\section{drift-gauge transformations of SDEs}
\label{sec:appD}

Remarkable feature of SDEs consists in the possibility to transform the deterministic (drift) terms keeping the averages to be unaffected by the transformation \cite{deuar_thesis_2005}. Mathematically, it means that we solve the new system of SDEs for individual trajectories, whereas describe the same physical system as before the transformation. However, in order to fulfill this requirement, one should also introduce the new stochastic variable called the weight function and update the averaging procedure. Let us recall the main ideas behind such a transformation.

Given the initial SDEs in the form
\begin{equation}
	\frac{d \vec{x} (t)}{dt} = \vec{\Mu} (\vec{x},t) + \vec{\xi} (\vec{x},t),
\end{equation}
\noindent one can change the deterministic part of the equation (which results in the new behavior of individual stochastic trajectories):
\begin{equation}
	\frac{d \vec{x}' (t)}{dt} = \vec{\Mu}' (\vec{x}',t) + \vec{\xi} (\vec{x}',t).
\label{eqnC:2}
\end{equation}

The averages of the stochastic variables remain unaffected by this transformation, if we also update the averaging procedure itself in the way as follows:
\begin{equation}
	\langle \mathcal{F} (\vec{x}(t)) \rangle = \langle \mathcal{F} (\vec{x}'(t)) \Omega(t) \rangle,
\label{eq:appD:newaverage}
\end{equation}
\noindent where $\Omega(t)= e^{C_0 (t)}$ is the time-dependent weight coefficient, and the evolution of $C_0(t)$ is governed by the following SDE:
\begin{equation}
	\frac{d C_0 (t)}{dt} = \Mu_0 (\vec{x}',t) + \xi_0 (\vec{x}',t).
\label{eq:appD:c0}
\end{equation}

The deterministic and stochastic terms in (\ref{eq:appD:c0}) should satisfy the following correlation properties (for details, see Chapter 4 of work \cite{deuar_thesis_2005}):
\begin{align}
	\langle \xi_0 (\vec{x}',t) \xi_0 (\vec{x}',t') \rangle &= - 2 \Mu_0 (\vec{x}',t) \delta(t-t'), \nonumber\\
	\langle \vec{\xi} (\vec{x}',t) \xi_0 (\vec{x}',t') \rangle &= - \bigl[\vec{\Mu}' (\vec{x}',t) - \vec{\Mu} (\vec{x}',t)\bigr] \delta (t-t').
\label{eq:appD:corr}
\end{align}

Following this idea, let us introduce such shifts of the deterministic terms of the field equations in (\ref{eqn:final}), that could lead to the emergence of an extra term $\sim w^4$ in (\ref{eqn:eff_potential}):
\begin{equation}
	\left(\frac{d \widetilde{\alpha}}{ d t}\right)_\text{det} = \left( \frac{d \alpha}{ d t} \right)_\text{det} + \Delta \alpha (t), \quad
		  \left(\frac{d \widetilde{\alpha}^\dagger }{ d t}\right)_\text{det} = \left( \frac{d \alpha^\dagger}{ d t} \right)_\text{det} + \Delta \alpha^\dagger  (t).
\label{eq:appD:aadnew}
\end{equation}

Introducing the variable $w(t) = 2 \rho_{ee}(t) -1$ and calculating its second-order derivative in time (here we consider the semi-classical case and omit the noise terms), we obtain:
\begin{equation}
	\ddot{w} = - f^2 \Bigl[ - 4 E w(t)  + 3 w(t)^2 - 1 \Bigr] + \frac{2 i f}{\sqrt{N}} \Bigl\{ \rho_{eg}(t) \Delta \alpha^\dagger  (t) - \rho_{ge}(t) \Delta \alpha  (t)  \Bigr\}.
\label{eq:appD:ddw}
\end{equation}

The first term in (\ref{eq:appD:ddw}) written in square brackets coincides with (\ref{eqn:anh_osc}) and leads to the effective potential (\ref{eqn:eff_potential}). Hence, the second term in (\ref{eq:appD:ddw}) can be defined from the condition:
\begin{equation}
	\frac{2 i f}{\sqrt{N}} \Bigl\{ \rho_{eg}(t) \Delta \alpha^\dagger  (t) - \rho_{ge}(t) \Delta \alpha  (t)  \Bigr\} \sim -f^2 w^3 (t),
\label{eq:appD:cond}
\end{equation}
\noindent resulting in the new term $\sim w^4$ after the formal integration of (\ref{eq:appD:ddw}) in time. In order to fulfill (\ref{eq:appD:cond}), we take into account the definition of the Bloch vector (\ref{eqn:e-bl}) and introduce $\Delta \alpha  (t)$ and $\Delta \alpha^\dagger  (t)$ in the form as follows:
\begin{gather}
	\Delta \alpha(t) = \varkappa \cdot i f\sqrt{N} \rho_{eg}(t) \rho_{ee}(t), \nonumber\\
	\Delta \alpha^\dagger(t) = - \varkappa \cdot i f\sqrt{N} \rho_{ge}(t) \rho_{ee}(t),
	\label{eqn:appD:a-shifts}
\end{gather}
\noindent where we also put an extra real-valued positive parameter $\varkappa$, which meaning is discussed in the main text. As a result, the effective potential (\ref{eqn:anh_osc}) undergoes the desired transformation:
\begin{equation}
	U(w) = \Bigl(2 E - \frac{\varkappa}{8} \Bigr)w^2 - \Bigl(1 - \frac{\varkappa}{8} \Bigr) w^3 + \Bigl( 1 + \frac{\varkappa}{8}\Bigr) w + \frac{\varkappa}{32} w^4.
	\label{eqn:appD:eff-pot-mod}
\end{equation}

Applying drift-gauge transformations (\ref{eqn:appD:a-shifts}) in SDEs (\ref{eqn:final}) should be accompanied by the inclusion of the stochastic equation for the variable $C_0(t)$. One should also satisfy the correlation properties between the $\widetilde{\alpha}(t)$, $\widetilde{\alpha}^\dagger (t)$ and $C_0(t)$:
\begin{gather}
	\langle \zeta(t) \xi_0 (t') \rangle = - \Delta \alpha(t) = - \varkappa \cdot i f\sqrt{N} \rho_{eg}(t) \rho_{ee}(t), \nonumber\\
	\langle \zeta^\dagger (t) \xi_0 (t') \rangle = - \Delta \alpha^\dagger(t) = \varkappa \cdot i f\sqrt{N} \rho_{ge}(t) \rho_{ee}(t),
\label{eq:appD:cor}
\end{gather}
\noindent where we use the same notations as in (\ref{eqnA1:new-noise}) or (\ref{gen:a}).

Taking into account the explicit form of the stochastic terms of the field equations in (\ref{eqn:final}), the correlators (\ref{eq:appD:cor}) can be fulfilled by means of the noises $S(t)$ (already present in the field equations) and $S^*(t)$ in the way as follows:
\begin{equation}
	\xi_0 (t) = \varkappa \cdot \sqrt{N} \rho_{ee}(t) S^*(t),
\end{equation}
\noindent which results in $\Mu_0 (t) = 0$. As a result, the SDE for the variable $C_0(t)$ takes the form:
\begin{equation}
	\frac{d C_0(t)}{dt} =  \varkappa \cdot \sqrt{N} \rho_{ee}(t) S^*(t).
\label{eqn:appD:eqC0}
\end{equation}

Empirically we also found out that better convergence is reached if one uses the real part of the stochastic variable $\rho_{ee} (t)$ in (\ref{eqn:appD:a-shifts}). This could be reasonable because we tend to modify the real-valued potential (\ref{eqn:eff_potential}) and bound the motion of the real part of $\rho_{ee}(t)$, when analysing its semi-classical behavior. Calculating the powers of complex numbers leads to the interference between their real and imaginary parts, that may result in totally unpredictable behaviour of (\ref{eqn:appD:eff-pot-mod}) when complex valued stochastic variable $\rho_{ee}(t)$ is used. Taking this finding into account and updating (\ref{eqn:appD:a-shifts}) and (\ref{eqn:appD:eqC0}), we derive (\ref{eqn:g-transform}).

\section{Stochastic conservation laws}
\label{sec:appD}

It is possible to show that certain initial conditions  can sometimes provide  conservation laws at the level of individual trajectories. As an example, let us consider (\ref{eqnA1:full}) and sum up the equations for the variables $\rho^{(\mu)}_{ee}(t)$ and $\rho^{(\mu)}_{gg}(t)$. As a result, we obtain:
\begin{equation}
	\frac{d}{d t} \biggl( \rho^{(\mu)}_{ee}(t) + \rho^{(\mu)}_{gg}(t) \biggr) = \Bigl( 1 - \bigl(\rho^{(\mu)}_{ee}(t) + \rho^{(\mu)}_{gg}(t)\bigr) \Bigr)
		\frac{f}{\sqrt{N}} \left[ \rho_{eg}^{(\mu)}(t) F^*(t) + \rho_{ge}^{(\mu)}(t) F^{\dagger *} (t)  \right].	
\end{equation} 
Provided that at the initial moment of time $\rho^{(\mu)}_{ee} (0) + \rho^{(\mu)}_{gg} (0) = 1$, we deduce
\begin{equation}
	\rho^{(\mu)}_{ee}(t) + \rho^{(\mu)}_{gg}(t) = 1, \quad \forall \, t > 0.
\label{eq:appE:iom1}	
\end{equation}

It is important to stress that conservation of a particular quantity also depends on the explicit form of the stochastic terms in SDEs (as we have freedom provided by the diffusion gauging of SDEs). As a particular example, one can mention the length of the stochastic Bloch vector 
\begin{equation}
	\Bigl(2\rho_{ee} (t) - 1 \Bigr)^2 + 4 \rho_{eg} (t) \rho_{ge} (t),
\end{equation}
\noindent which in the case of the closed TCM ($\gamma = 0$) is conserved (equals unity, if we start with the appropriate initial conditions) within the equations (\ref{eqnA3:simpl01}), but is not the integral of motion within (\ref{eqn:final}).

\section{Additional remarks on the noise terms}
\label{sec:appE}

In this section we perform qualitative analysis of the contributions from the noise terms by comparing the explicit form of quantum and averaged stochastic equations. For this purpose we consider the equation for the excited state probability (and its stochastic counterpart) in the case of a closed TCM ($\gamma = 0$) and exact resonance $\omega_c = \omega_0$. We start with the observable 
\begin{equation}
	\langle \hat{\sigma}_{ee} \rangle  = \Tr{\Bigl(\hat{\sigma}_{ee} \hat{\rho} (t) \Bigr)}, \quad \hat{\sigma}_{ee} = \frac{1}{N} \sum_{\mu = 1}^{N} \hat{\sigma}^{(\mu)}_{ee}. 
\label{eq:appE:sigma}
\end{equation}

Differentiating (\ref{eq:appE:sigma}) once and twice over time, we obtain:
\begin{eqnarray}
	\frac{d \langle \hat{\sigma}_{ee} \rangle}{dt}  = \Tr{\Bigl(\hat{\sigma}_{ee} \dot{\hat{\rho}} (t) \Bigr)} = \frac{i}{\hbar} \Tr{\Bigl(\hat{\sigma}_{ee} [\hat{\rho}(t), \hat{H}] \Bigr)} =
		\frac{i f}{\sqrt{N}} \frac{1}{N} \sum_{\mu = 1}^{N} \Tr{ \Bigl( (\hat{\sigma}_{ge}^{(\mu)} \hat{a} - \hat{\sigma}_{eg}^{(\mu)} \hat{a}^\dagger) \hat{\rho} (t)  \Bigr)}, \nonumber\\
	\frac{d^2 \langle \hat{\sigma}_{ee} \rangle}{dt^2} = \frac{i f}{\sqrt{N}} \frac{1}{N} \sum_{\mu = 1}^{N} \Tr{ \Bigl( (\hat{\sigma}_{ge}^{(\mu)} \hat{a} - \hat{\sigma}_{eg}^{(\mu)} \hat{a}^\dagger) \dot{\hat{\rho}} (t)  \Bigr)} =
		\frac{i f}{\sqrt{N}} \frac{1}{N} \frac{i}{\hbar} \sum_{\mu = 1}^{N} \Tr{ \Bigl( [\hat{H}, (\hat{\sigma}_{ge}^{(\mu)} \hat{a} - \hat{\sigma}_{eg}^{(\mu)} \hat{a}^\dagger)] \hat{\rho} (t)  \Bigr)} = \nonumber\\
	- \frac{f^2}{N^2} \sum_{\mu = 1}^{N} \Tr{ \Bigl(  2 \hat{\sigma}^{(\mu)}_{ee} + 2 (\hat{\sigma}^{(\mu)}_{ee}-\hat{\sigma}^{(\mu)}_{gg}) \hat{a}^\dagger \hat{a} +
			\sum_{\nu \neq \mu} ( \hat{\sigma}_{eg}^{(\nu)} \hat{\sigma}_{ge}^{(\mu)} + \hat{\sigma}_{ge}^{(\nu)} \hat{\sigma}_{eg}^{(\mu)} 	 ) \hat{\rho} (t) \biggr)}.
\label{eq:appE:ddotP}
\end{eqnarray}
Taking the ansatz (\ref{eqnA1:rho-tot})--(\ref{eqnA1:rho-at}) for the density operator, we derive:
\begin{equation}
	\frac{d^2 \rho_{ee}(t)}{dt^2} = - \frac{2}{f^2} \biggl[ \frac{1}{N} \bigl(2 \rho_{ee}(t) - 1\bigr) \alpha^\dagger(t) \alpha(t) + \frac{1}{N} \rho_{ee}(t) + \rho_{eg}(t) \rho_{ge}(t) \left(1-\frac{1}{N} \right)   \biggr],
\end{equation}
\noindent where we took into account the symmetry of the atomic subsystem $\rho_{ij}^{(\mu)} (t) \rightarrow \rho_{ij} (t)$.

Now let us take the average of the stochastic equation (\ref{eqn:final}) for the variable $\rho_{ee} (t)$ (here we take into account that equations (\ref{eqn:final}) are written in the Ito's form):
\begin{equation}
	\left\langle \frac{d \rho_{ee}(t)}{ d t} \right\rangle = i \frac{f}{\sqrt{N}} \Bigl( \langle \rho_{eg}(t) \alpha^\dagger(t) \rangle - \langle \rho_{ge}(t) \alpha(t) \rangle \Bigr).
\label{eq:appE:pee_1}
\end{equation}
Differentiating (\ref{eq:appE:pee_1}) over the time and using the Ito's lemma \cite{ito_stochastic_1944}, we obtain:
\begin{eqnarray}
	\frac{d}{dt} \left\langle \frac{d \rho_{ee}(t)}{ d t} \right\rangle = - \frac{2}{f^2} \Bigl[ 
		\frac{1}{N}\langle \bigl( 2 \rho_{ee} (t) - 1 \bigr) \alpha^\dagger(t) \alpha(t)   \rangle +
		\frac{1}{N} \langle \rho_{ee} (t) \rangle \frac{ \mathbb{1}_{F} + \mathbb{1}_{F^\dagger} }{2} +
		\langle \rho_{eg}(t) \rho_{ge}(t) \rangle
		\left( 1 - \frac{1}{N} \cdot \mathbb{1}_S \right)\Bigr],
\label{eq:appE:avstoch}
\end{eqnarray}
\noindent where we introduced the notations $\mathbb{1}_{F}$, $\mathbb{1}_{F^\dagger}$, $\mathbb{1}_S$ for the unity values  to explicitly point out their origin from the associated correlators:
\begin{equation}
	\langle F (t) F^{*} (t') \rangle = \mathbb{1}_{F} \cdot \delta(t- t'), \quad 
		\langle F^\dagger (t) F^{\dagger *} (t') \rangle = \mathbb{1}_{F^\dagger} \cdot	\delta(t- t'), \quad
		\langle S (t) S^* (t') \rangle = \mathbb{1}_S \cdot \delta(t- t').
\end{equation}

Let us stress that the terms containing $1/N$ factor emerge in (\ref{eq:appE:ddotP}) due to the non-zero commutators between the operators $\hat{\sigma}_{ee}^{(\mu)}$, $\hat{\sigma}_{eg}^{(\mu)}$ and $\hat{\sigma}_{ge}^{(\mu)}$. Comparing (\ref{eq:appE:avstoch}) and (\ref{eq:appE:ddotP}), we see that sampled solutions of the stochastic equations reproduce the corresponding commutators, defined by the properties of quantum operators.  

\end{document}